\pdfoutput=1

\documentclass[aps,twocolumn,amsmath,amssymb,preprintnumbers,floatfix,prb,longbibliography]{revtex4-2}
\usepackage{times}
\usepackage{amsfonts}
\usepackage{amsmath, amsthm, amssymb}
\usepackage{color}
\usepackage{bm}
\usepackage{natbib}
\usepackage[caption=false]{subfig}
\usepackage[colorlinks=True, allcolors=blue]{hyperref}
\usepackage{siunitx}
\usepackage[capitalise]{cleveref}
\usepackage[flushleft]{threeparttable}

\usepackage{graphicx}

\newcommand{\cf}{cf.~}
\newcommand{\ie}{i.e.~}
\newcommand{\eg}{e.g.~}

\renewcommand{\vec}[1]{\mathbf{#1}}
\newcommand{\vq}{\vec{q}}
\newcommand{\vp}{\vec{p}}
\newcommand{\vk}{\vec{k}}

\newcommand{\On}{\Omega_n}
\newcommand{\Om}{\Omega_m}
\newcommand{\on}{\omega_n}

\renewcommand{\P}{D}
\newcommand{\p}{d}

\renewcommand{\sd}{\downarrow}
\newcommand{\su}{\uparrow}

\renewcommand{\l}{\left}
\renewcommand{\r}{\right}

\usepackage{soul}

\allowdisplaybreaks

\begin{document}
\title{Cavity-mediated superconductor--ferromagnetic insulator coupling}

\author{Andreas T. G. Janss{\o}nn}
\author{Henning G. Hugdal}
\email[Corresponding author: ]{henning.g.hugdal@ntnu.no}
\author{Arne Brataas}
\author{Sol H. Jacobsen}

\affiliation{Center for Quantum Spintronics, Department of Physics, NTNU Norwegian University of Science and Technology, NO-7491 Trondheim, Norway}

\begin{abstract}
A recent proof of concept showed that cavity photons can mediate superconducting (SC) signatures to a ferromagnetic insulator (FI) over a macroscopic distance [Phys. Rev. B, \textbf{102,} 180506(R) (2020)]. In contrast with conventional proximity systems, this facilitates long-distance FI--SC coupling, local subjection to different drives and temperatures, and studies of their mutual interactions without proximal disruption of their orders. Here we derive a microscopic theory for these interactions, with an emphasis on the leading effect on the FI, namely, an induced anisotropy field. In an arbitrary practical example, we find an anisotropy field of $14$--$\SI{16}{\micro T}$, which is expected to yield an experimentally appreciable tilt of the FI spins for low-coercivity FIs such as Bi-YIG. We discuss the implications and potential applications of such a system in the context of superconducting spintronics.
\end{abstract}
\maketitle

\section{Introduction}\label{Sec:Intro}

Enabling low-dissipation charge and spin transport, superconducting spintronics presents a pathway to reducing energy costs of data processing, and provides fertile ground for exploring new fundamental physics~\cite{Eschrig2011,Linder2015,Joshi2016}. Conventionally, superconducting and spintronic systems are coupled by the proximity effect, with properties of adjacent materials transported across an interface. The superconducting coherence length thus limits the extent to which superconducting properties can be harnessed in proximity systems, to a range of nm--$\mu$m near interfaces~\cite{Tokuyasu1988, Demler1997,Keizer2006,Anwar2010,Eschrig2015}.

By contrast, cavity-coupled systems offer mediation across macroscopic distances~\cite{Johansen2018, Janssonn2020, Tabuchi2014, Tabuchi2015, Tabuchi2016}. They also offer interaction strengths that relate inversely to the cavity volume~\cite{Schlawin2019, Kakazu1994}, which is routinely utilized experimentally to achieve strong coupling in e.g. GHz--THz cavity set-ups~\cite{Liu2015, Zhang2016, Bayer2017, Keller2017, Schlawin2022}. Furthermore, research on the coupling of magnets and cavity photons shows that the effective interaction strengths scale with the number of spins involved~\cite{Johansen2018, Johansen2019, Soykal2010, Soykal2010b, Liu2016, ViolaKusminskiy2016, Schlawin2022, Harder2018, Rameshti2022, Yuan2022}, which has been utilized experimentally to achieve effective coupling strengths far exceeding losses~\cite{Huebl2013, Bourhill2016, Tabuchi2014, Tabuchi2016, Bai2015, LachanceQuirion2019, Liensberger2021, Khan2021, Abdurakhimov2019, Zhang2014, Schlawin2022, Harder2018, Rameshti2022, Yuan2022}.

Theoretically, a number of methods have been employed to extract mediated effects in cavity-coupled systems. This includes, but is not limited to, classical modelling for coupling two ferromagnets~\cite{Rameshti2018}, and a ferromagnet to a superconductor~\cite{Janssonn2020}; application of Jaynes--Cummings-like models for coupling a ferromagnet and a qubit~\cite{Tabuchi2015, Tabuchi2016, LachanceQuirion2019, LachanceQuirion2017}, and two ferromagnets~\cite{Lambert2016}; perturbative diagonalization by the Schrieffer--Wolff transformation for coupling a ferro- and antiferromagnet~\cite{Johansen2018, Johansen2019, Yuan2022}, and a normal metal to itself~\cite{Schlawin2019, Schlawin2022}; and perturbative evolution of the density matrix, as well as perturbative diagonalization by the non-equilibrium Keldysh path integral formalism, for coupling a mesoscopic circuit to a cavity~\cite{Cottet2020}.

\begin{figure}[t]
    \centering
    \includegraphics[width=1\linewidth]{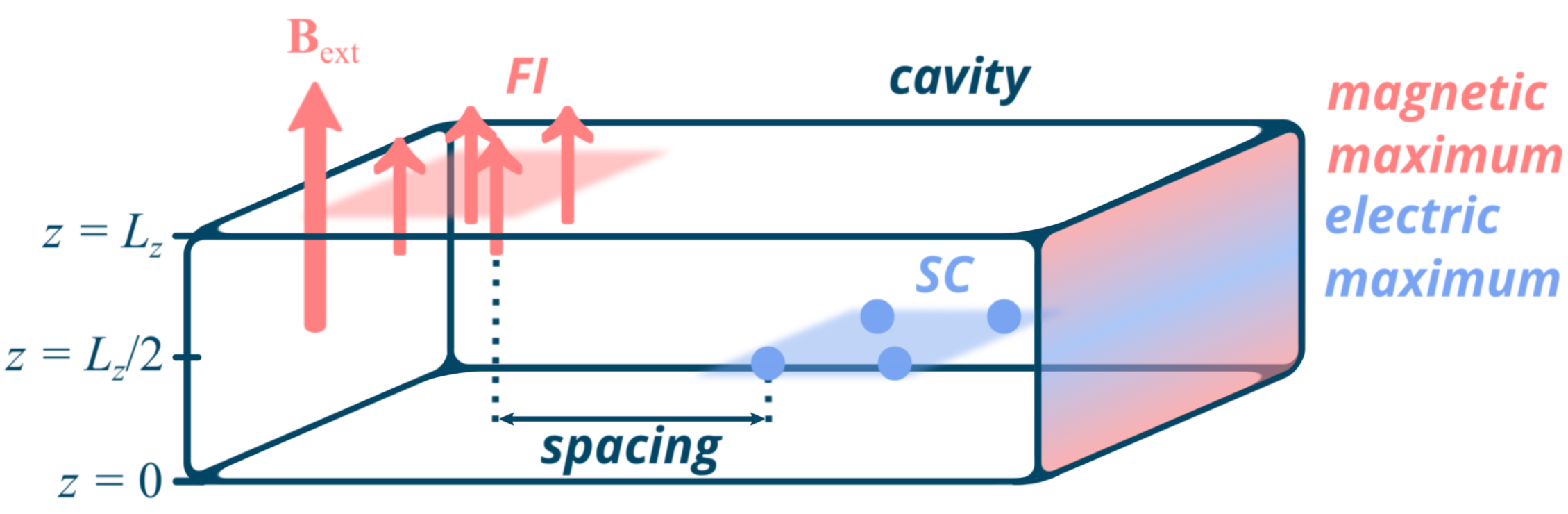}
    \caption{Illustration of the set-up. A thin ferromagnetic insulator and thin superconductor are placed spaced apart inside a rectangular, electromagnetic cavity. The FI is subjected to an aligning external magnetic field $\vec{B}_{\mathrm{ext}}$. The cavity is short along the $z$ direction, and long along the perpendicular $xy$ directions, causing cavity modes to separate into a band-like structure. The FI and the SC are respectively placed in regions of maximum magnetic ($z = L_z$) and electric ($z = L_z/2$) cavity field of the $\ell_z = 1$ modes, as defined in Sec.~\ref{sec:CavityGaugeField} and illustrated above by the colored field cross-section on the right wall.}
    \label{fig:set-up}
\end{figure}

In this paper, we will employ the Matsubara path integral formalism~\cite{Altland, Kopnin, Kachelriess, Negele, Roman-Roche2022} to derive a microscopic theory for the cavity-mediated coupling of a ferromagnetic insulator (FI) with a singlet $s$-wave superconductor (SC). In particular, we consider the Zeeman coupling to the FI, and the paramagnetic coupling to the SC. We show that with this approach, we may exactly integrate out the net mediated effect by the cavity photons. This is in contrast to the Schrieffer--Wolff approach, which would limit the integrating-out of the cavity to off-resonant regimes~\cite{Johansen2019}. For instance, a pairing term analogous to the one found via the Schrieffer--Wolff transformation in Ref.~\cite{Schlawin2019} also appears in our calculations, without the limitation to an off-resonant regime. Furthermore, unlike many preceding works which single out the coupling to the uniform mode of the magnet~\cite{Janssonn2020, Johansen2018, Bourhill2016, Tabuchi2016, Huebl2013, Soykal2010}, we retain the influence of a range of modes in our model. Their non-negligible influence when the magnet exceeds a certain size relative to the cavity, has been emphasized by both experimentalists \cite{Bourhill2016} and theorists \cite{Soykal2010}.

The Matsubara path integral approach was very recently applied to construct a general effective theory of cavity-coupled material systems of identical particles~\cite{Roman-Roche2022}, highlighting some of the same advantages of this approach as above. By contrast, we consider the cavity-mediated coupling of lattices of two distinct classes of quasiparticles, specifically magnons and SC quasiparticles.

By a careful choice of cavity dimensions and the placement of subsystems, we couple the insulator to the momentum degrees of freedom of the superconductor. In this case, the cavity acts as an effective spin--orbit coupling. Here, we emphasize the leading effect of the superconductor on the insulator, namely, the induction of an anisotropy field. In an arbitrary, practical example, we achieve a field of $14$--$\SI{16}{\micro T}$, which is expected to yield an experimentally appreciable tilt of the FI spins for an insulator of sufficiently low coercivity such as Bi-YIG. Since the cavity facilitates coupling across unconventionally long distances, it enables the FI and SC to be held at different temperatures, be subjected separately to external drives, and have them interact without the same mutual disruption of their orders associated with the proximity effect \cite{Janssonn2020, Linder2015}, such as the breaking of Cooper pairs by magnetic fields from the FI. In practical applications, our system may be used to bridge superconducting and other spintronic circuitry.

The article is organised as follows. In~\cref{sec:set-up} we present the set-up: A cavity with an FI and SC film placed at magnetic and electric antinodes as shown in Fig.~\ref{fig:set-up}, with no overlap in the $xy$ plane. In~\cref{sec:Hamiltonian} we cover theoretical preliminaries: The quantized gauge field, the magnon-basis Hamiltonian for the insulator, and the Bogoliubov quasiparticle-basis Hamiltonian for the superconductor. The system Hamiltonian is subsequently constructed. In Sec.~\ref{sec:Matsubara}--\ref{sec:IntOutSC}, we construct an effective magnon theory using the path integral formalism. Here we exactly integrate out the cavity, and perturbatively the superconductor. In~\cref{subsec:result_electric}, we extract from the effective theory the leading effect of the superconductor on the insulator, namely, the induced anisotropy field. In a practical example, we calculate this field numerically, and find here an induced field on the order of \si{\micro\tesla} in magnitude. Finally, in~\cref{sec:conclusion}, we give concluding remarks, discussing the results and their significance, and an outlook. In the appendices, we affirm the mathematical consistency of the effective theory with an alternative derivation, explore a variation of the set-up with the SC placed at the opposite magnetic antinode, and elaborate on the interpretation of certain quantities in the effective action as an effective anisotropy field.

\section{Theory}\label{Sec:Theory}

\subsection{Set-up}\label{sec:set-up}

Our set-up is illustrated in Fig.~\ref{fig:set-up}. We place two thin layers, one of a ferromagnetic insulator (FI) and one of a superconductor (SC), spaced apart inside a rectangular electromagnetic cavity. The dimensions of the cavity are $L_x, L_y \gg L_z$, with $L_z$ on the $\mu$m--mm scale, and $L_x, L_y$ on the cm scale. The aspect ratios render photons more easily excited in the $xy$ directions. The FI is placed at the upper magnetic antinode of the $\ell_z = 1$ modes (\cf Sec.~\ref{sec:CavityGaugeField}), and the SC at the corresponding electric antinode, as illustrated in Fig.~\ref{fig:set-up}. Because the layers are thin in comparison to $L_z$, the local spatial variation of the modes in the $z$ direction is negligible, i.e., the modes are treated as uniform in the $z$ direction.

The FI is locally subjected to an aligning and perpendicular uniform, external magnetostatic field, which vanishes across the SC. This was achieved experimentally with external coils and magnetic shielding in \citet{Tabuchi2016}. Furthermore, the SC is subjected to an in-plane supercurrent. This may be realized by passing a direct current (DC) through small electric wires, entering the cavity via small holes in the walls and connecting along the sides of the SC, similarly to Ref.~\cite{Takasan}. Provided the wires and holes are sufficiently small, their influence on the cavity modes are negligible. Provided the sample width does not exceed the Pearl length $\lambda^2/d_{\mathrm{SC}}$~\cite{Pearl1964,Tinkham, Takasan}, the leading effect of the DC is to induce an equilibrium supercurrent with a Cooper pair center-of-mass momentum $2\vec{P}$, with the magnitude of $\vec{P}$ determined by the current. Here $\lambda$ is the effective magnetic penetration depth, and $d_{\mathrm{SC}}$ is the sample depth. For Nb thin films, we expect the Pearl length criterion to be met at widths of up to $\SI{0.1}{mm}$ for a $d_{\mathrm{SC}}$ down to $\SI{1}{nm}$~\cite{Gubin2005}. 

\subsection{Hamiltonian} \label{sec:Hamiltonian}

In the following, we deduce a Hamiltonian
\begin{equation}
    \mathcal{H} \equiv \mathcal{H}_{\mathrm{FI}} + \mathcal{H}^{\mathrm{cav}}_0 + \mathcal{H}_{\mathrm{SC}}.
\end{equation}
for the system illustrated in Fig.~\ref{fig:set-up}. We begin by quantizing the cavity gauge field, and introducing the cavity Hamiltonian $\mathcal{H}^{\mathrm{cav}}_0$. Following this, we deduce a Hamiltonian $\mathcal{H}_{\mathrm{FI}}$ for the FI in the magnon basis, including the Zeeman coupling to the cavity. Finally, we deduce a Hamiltonian $\mathcal{H}_{\mathrm{SC}}$ for the SC in the quasiparticle basis, including the paramagnetic coupling to the cavity.

\subsubsection{Cavity gauge field} \label{sec:CavityGaugeField}

We begin by presenting the expression for the quantized cavity gauge field $\vec{A}_{\mathrm{cav}}$~\cite{Kakazu1994}. Starting from the Fourier decomposition of the classical vector potential, we impose the transverse gauge and quantize the field. We employ reflecting boundary conditions at the cavity walls in the $z$ direction, and periodic boundary conditions at the comparatively distant walls in the $xy$ directions. The gauge field is thus
\begin{equation} \label{eq:Acavgauged}
    \vec{A}_{\mathrm{cav}} \equiv \sum_{\vec{Q}  \varsigma} \sqrt{\frac{\hbar}{2\epsilon \omega_{\vec{Q}}}} ( a_{\vec{Q}  \varsigma} \vec{\bar{u}}_{\vec{Q}  \varsigma} + a_{\vec{Q}  \varsigma}^\dagger \vec{\bar{u}}_{\vec{Q}  \varsigma}^*).
\end{equation}
Above,
\begin{equation} \label{eq:Qvec}
\begin{split}
    \vec{Q} & \equiv (Q_x, Q_y, Q_z) \equiv (2\pi \ell_x/L_x, 2\pi \ell_y/L_y, \pi \ell_z/L_z)
\end{split}
\end{equation}
are the momenta of each photonic mode, with $\ell_x,\ell_y = 0, \pm 1, \pm 2, \dots$ and $\ell_z = 0,1,2, \dots$. The discretization of $Q_z$ differs from that of $Q_x$ and $Q_y$ due to the different boundary conditions in the transverse and longitudinal directions. Furthermore, $\varsigma = 1,2$ labels polarization directions, $\epsilon$ is the permittivity of the material filling the cavity, and 
\begin{equation}
    \omega_{\vec{Q}} = c|\vec{Q}|
\end{equation}
is the cavity dispersion relation, with $c$ the speed of light. $a_{\vec{Q} \varsigma}^\dagger$ and $a_{\vec{Q} \varsigma}$ are photon creation and annihilation operators, satisfying
\begin{equation}
    [a_{\vec{Q} \varsigma},a_{\vec{Q}^\prime \varsigma^\prime}^\dagger] = \delta_{\vec{Q} \vec{Q}^\prime} \delta_{\varsigma \varsigma^\prime},
\end{equation}
where the factors on the right-hand side are Kronecker delta functions.

\begin{figure}[ht]
    \centering
    \includegraphics[width=.5\linewidth]{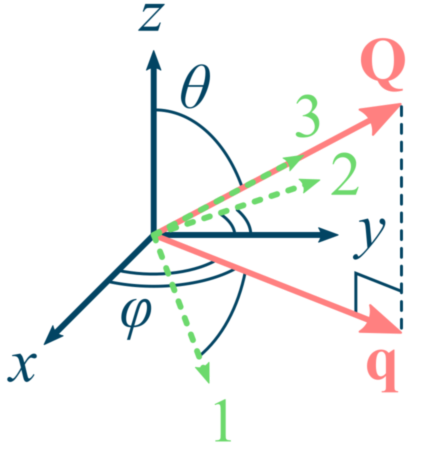}
    \caption{Illustration of the $123$ coordinate system. $\vec{Q}$ is the photon momentum vector, and $\vec{q}$ is its component in the $xy$ plane. $\theta$ (single line) is the polar, and $\varphi$ (double line) the azimuthal angle associated with $\vec{Q}$ in relation to the $xyz$ basis. The $123$ axes results from a rotation of the $xyz$ axes by an angle $\theta$ about the $y$ axis, followed by a rotation by an angle $\varphi$ about the original $z$ axis. In the illustration, the $1$ axis points somewhat outwards and downwards, the $2$ axis points somewhat inwards and is confined to the original $xy$ plane, and the $3$ axis aligns with $\vec{Q}$.}
    \label{fig:123Q}
\end{figure}

Lastly, the mode functions
\begin{equation} \label{eq:123modes}
    \vec{\bar{u}}_{\vec{Q}  \varsigma} \equiv \sum_{\P} \hat{e}_{\P} O^{\vec{Q}}_{ \varsigma \P} u_{\vec{Q} \P}
\end{equation}
encapsulate the spatial modulation of the modes. Here, $\hat{e}_{\P}$ is the unit vector in the $\P = x,y,z$ direction. $O^{\vec{Q}}_{ \varsigma \P}$ are elements of a matrix that rotates the original $xyz$ basis of unit vectors to a new basis labeled $123$, with the $3$ direction aligned with $\vec{Q}$ (see Fig.~\ref{fig:123Q}):
\begin{equation} \label{eq:123basis}
    \begin{pmatrix}
    \hat{e}^{\vec{Q}}_1 \\
    \hat{e}^{\vec{Q}}_2 \\
    \hat{e}^{\vec{Q}}_3
    \end{pmatrix}
    = O^{\vec{Q}} 
    \begin{pmatrix}
    \hat{e}_x \\
    \hat{e}_y \\
    \hat{e}_z
    \end{pmatrix},
\end{equation}
\begin{equation}
    O^{\vec{Q}} \equiv
    \begin{pmatrix}
    \cos \theta \cos \varphi & \cos \theta \sin \varphi & -\sin \theta \\
    -\sin \varphi & \cos \varphi & 0 \\
    \sin \theta \cos \varphi & \sin \theta \sin \varphi & \cos \theta
    \end{pmatrix}.
\end{equation}
Here $\theta = \theta_{\vec{Q}}$ and $\varphi = \varphi_{\vec{Q}}$ are the polar and azimuthal angles illustrated in Fig.~\ref{fig:123Q}. $O^{\vec{Q}}$ originates from the implementation of the transverse gauge, which amounts to neglecting the longitudinal $3$ component of the gauge field. Finally, $u_{\vec{Q} \P}$ are the mode functions in the $xyz$ basis, given by
\begin{align} \label{eq:uxyz}
    u_{\vec{Q} x} & = u_{\vec{Q} y} = \sqrt{\frac{2}{V}} e^{iQ_x x + iQ_y y} i\sin Q_z z, \\
    u_{\vec{Q} z} & = \sqrt{\frac{2}{V}} e^{iQ_x x + iQ_y y} \cos Q_z z,
\end{align}
where $V$ is the volume of the cavity~\footnote{We have neglected a prefactor $\sqrt{w_{\ell_z} w_{z}}$ of $u_{\vec{Q} z}$ which is inconsequential to us. The function $w_x = 1/2$ when $x = 0$, and $w_x = 1$ otherwise. This prefactor follows from the reflecting boundary conditions combined with requiring that the Fourier transformation be unitary.}.

Our set-up facilitates coupling to the $\ell_z = 1$ band of cavity modes, as the FI and SC are placed in field maxima as illustrated in Fig.~\ref{fig:set-up}. We will only consider variations of the in-plane part $\vq$ of the general momenta $\vec{Q}$, defined via
\begin{equation} \label{eq:Qtoq}
    \vec{Q} \equiv \vec{q} + \pi \hat{e}_z/L_z.
\end{equation}
For this reason we will use the subscript $\vq$ for functions of $\vec{Q}$ where the $z$ component is locked to the $\ell_z = 1$ mode, \eg
\begin{equation} \label{eq:cavitydispersion}
    \omega_{\vec{q}} \equiv \omega_{\vec{Q}} \big|_{\vec{Q} = \vec{q} + \pi \hat{e}_z/L_z} = c \sqrt{\left( \frac{\pi}{L_z} \right)^2 + \vec{q}^2}.
\end{equation}

The cavity itself contributes to the system Hamiltonian with the term
\begin{equation}
    \mathcal{H}^{\mathrm{cav}}_0 \equiv \sum_{\vec{q} \varsigma} \hbar \omega_{\vec{q}} a_{\vec{q} \varsigma}^\dagger a_{\vec{q} \varsigma},
\end{equation}
where we have disregarded the zero-point energy, since it does not influence our results.

\subsubsection{Ferromagnetic insulator}

The Hamiltonian of the FI in the cavity is
\begin{align}
    \mathcal{H}_{\mathrm{FI}} & \equiv \mathcal{H}_{\mathrm{ex}} + \mathcal{H}_{\mathrm{ext}} + \mathcal{H}_{\mathrm{FI-cav}}, \label{eq:Hspin}
\end{align}
with
\begin{subequations}
\begin{align}
    \mathcal{H}_{\mathrm{ex}} & \equiv -J \sum_{\langle i,j \rangle} \vec{S}_i \cdot \vec{S}_j, \\
    \mathcal{H}_{\mathrm{ext}} & \equiv
    - \frac{g\mu_B}{\hbar} B_{\mathrm{ext}} \sum_i S_{iz}, \\
    \mathcal{H}_{\mathrm{FI-cav}} & \equiv - \frac{g\mu_B}{\hbar}\sum_i \vec{S}_{i} \cdot \vec{B}_{\mathrm{cav}}(\vec{r}_i). \label{eq:HFICavSpinBasis}
\end{align}
\end{subequations}
The first term is the exchange interaction: $J > 0$ is the exchange interaction strength for a ferromagnetic insulator, $\vec{S}_i$ is the spin at lattice site $i$, and only nearest neighbor interactions are taken into account, as indicated by the angle brackets. The next two terms are Zeeman couplings: $g$ is the gyromagnetic ratio, $\mu_B$ is the Bohr magneton, $B_{\mathrm{ext}}$ is a strong (\ie $|B_{\mathrm{ext}}| \gg |\vec{B}_{\mathrm{cav}}|$) and uniform external magnetostatic field aligning the spins in the $z$ direction, and $\vec{B}_{\mathrm{cav}}(\vec{r}_i)$ is the magnetic component of the cavity field at lattice site $i$. The corresponding position vector is $\vec{r}_i$.

It is convenient to transition from the spin basis $\{ S_{ix}, S_{iy}, S_{iz} \}$ to a bosonic magnon basis $\{ \eta_i, \eta_i^\dagger \}$. This is achieved with the Holstein--Primakoff transformation~\cite{Holstein1940}, which is covered in detail in Refs.~\cite{Kittel, Johansen2019}.

Each FI lattice site carries spin $S$. The aligning field $B_{\mathrm{ext}}$ regulates the excitation energy of magnons (\cf Eq.~\eqref{eq:magnondispersion}), hence a sufficiently strong field implies few magnons per lattice site, \ie
\begin{equation} \label{eq:assumeFewMagnons}
    \langle \eta_i^\dagger \eta_i \rangle \ll 2S.
\end{equation}
We can therefore Taylor-expand the Holstein--Primakoff transformation, leading to the relations
\begin{equation} \label{Siz2}
    S_{iz} = \hbar(S - \eta_i^\dagger \eta_i),
\end{equation}
\begin{equation} \label{eq:Sipmsimplecartesian2}
    S_{i\p} \approx \frac{\hbar\sqrt{2S}}{2} (\nu_{\p} \eta_i + \nu_{\p}^* \eta_i^\dagger),
\end{equation}
where $\p = x,y$ and $\{ \nu_x, \nu_y \} = \{ 1, -i \}$.

Now, upon Fourier-decomposing the magnon operators
\begin{equation} \label{eq:magnFourierInvPosMom}
    \begin{split}
        \eta_{\vec{r}_i} & \equiv \frac{1}{\sqrt{N_{\mathrm{FI}}}} \sum_{\vec{k}} \eta_{\vec{k}} e^{i \vec{k} \cdot \vec{r}_i},
    \end{split}
\end{equation}
we obtain the conventional expression for $\mathcal{H}_{\mathrm{ex}} + \mathcal{H}_{\mathrm{ext}}$ in the magnon basis~\cite{Kittel}:
\begin{equation} \label{eq:diagMagnon}
    \mathcal{H}_{\mathrm{ex}} + \mathcal{H}_{\mathrm{ext}} \approx \mathcal{H}_0^{\mathrm{FI}} \equiv \sum_{\vec{k}} \hbar \lambda_{\vec{k}} \eta_{\vec{k}}^\dagger \eta_{\vec{k}},
\end{equation}
where we have introduced the magnon dispersion relation
\begin{equation} \label{eq:magnondispersion}
    \lambda_{\vec{k}} \equiv 2\hbar JN_{\delta}S \left(1 - \frac{1}{N_{\delta}} \sum_{\bm{\delta}} e^{i\vec{k}\cdot\bm{\delta}} \right) + \frac{g\mu_B}{\hbar} B_{\mathrm{ext}}.
\end{equation}
Above, $N_{\mathrm{FI}}$ is the total number of FI lattice points, $N_{\delta} = 6$ is the number of nearest-neighbor lattice sites on a cubic lattice (neglecting edges and corners), and $\bm{\delta} = \pm a_{\mathrm{FI}} \hat{e}_x, \pm a_{\mathrm{FI}} \hat{e}_y, \pm a_{\mathrm{FI}} \hat{e}_z$ are nearest-neighbor lattice vectors. The magnon momenta are
\begin{align}
    \vec{k} & \equiv (2\pi m_x^{\mathrm{FI}} / l_x^{\mathrm{FI}}, 2\pi m_y^{\mathrm{FI}} / l_y^{\mathrm{FI}}, 0) \equiv (k_x, k_y, 0), \label{eq:MomFI}
\end{align}
where $m_{\p}^{\mathrm{FI}} = -\left\lfloor \frac{N_{\p}^{\mathrm{FI}} - 1}{2} \right\rfloor, \dots, N_{\p}^{\mathrm{FI}} - 1 -\left\lfloor \frac{N_{\p}^{\mathrm{FI}} - 1}{2} \right\rfloor$ covers the first Brillouin zone (1BZ), with $N_{\p}^{\mathrm{FI}}$ the number of FI lattice points in direction $\p$, and $\lfloor \cdot \rfloor$ the floor function. Here we neglect the $k_z$ component; only the $k_z = 0$ modes enter our calculations due to the thinness of the FI film (cf. Eq.~\eqref{eq:DsumFI}). Note that the set of magnon momenta generally does not overlap with that of photon momenta in \cref{eq:Qvec}. Observe furthermore that the magnon energies~\eqref{eq:magnondispersion} can easily be regulated experimentally by adjusting $B_{\mathrm{ext}}$.

Proceeding to the interaction term, we deduce the magnetic cavity field $\vec{B}_{\mathrm{cav}}(\vec{r}_i)$ across the FI, which is the curl of the gauge field at $z \approx L_z$:
\begin{equation} \label{eq:Bcav}
\begin{split}
    & \vec{B}_{\mathrm{cav}}(\vec{r}_i) \big|_{\mathrm{FI}} = \nabla \times \vec{A}_{\mathrm{cav}}(\vec{r}_i) \big|_{\mathrm{FI}} \\
    = & - \sum_{\vec{q} \p} i \nu_{\p}^2 q_{\bar{\p}} \hat{e}_{\p} \sin \theta_{\vec{q}} \sqrt{\frac{\hbar}{\epsilon \omega_{\vec{q}}V}} e^{i\vec{q}\cdot\vec{r}_i} \cos\frac{\pi z_i}{L_z} ( a_{\vec{q}  1} + a_{-\vec{q}  1}^\dagger).
\end{split}
\end{equation}
Above, $\bar{\p}$ ``inverts'' $\p$ such that $\bar{x} = y$ and $\bar{y} = x$, and $z_i$ is the $z$ position of lattice site $i$. Note that the photon momentum component $q_{\bar{\p}}$ enters the sum with an inverted lower index. Observe that only the $1$ direction enters the expression, because $\vec{A}_{\mathrm{cav}}$ at $z \approx L_z$ points purely along the $z$ direction. The $2$ direction is by definition locked to the $xy$ plane, and does therefore not contribute at $z \approx L_z$.

Inserting Eqs.~\eqref{Siz2}--\eqref{eq:magnFourierInvPosMom} and \eqref{eq:Bcav} into Eq.~\eqref{eq:HFICavSpinBasis}, we find
\begin{equation} \label{eq:magnoninteraction2}
\begin{split}
    \mathcal{H}_{\mathrm{FI-cav}} & \approx \sum_{\vec{k} \p} \sum_{\vec{q} \varsigma} g_{\p}^{\vec{k} \vec{q}} ( \nu_{\p} \eta_{-\vec{k}} + \nu_{\p}^* \eta_{\vec{k}}^\dagger) ( a_{\vec{q} 1} + a_{-\vec{q} 1}^\dagger ),
\end{split}
\end{equation}
and hence a complete FI Hamiltonian $\mathcal{H}_{\mathrm{FI}} \approx H_0^{\mathrm{FI}} + \mathcal{H}_{\mathrm{FI-cav}}$. Above, we defined the coupling strength
\begin{equation} \label{eq:magnoncoupling2}
    g_{\p}^{\vec{k} \vec{q}} \equiv -g\mu_B q_{\bar{\p}} i\nu_{\p}^2 \sin \theta_{\vec{q}} \sqrt{\frac{S \hbar N_{\mathrm{FI}}}{2\epsilon \omega_{\vec{q}}V}} D_{\vec{k}\vec{q}}^{\mathrm{FI}} e^{i \vec{q} \cdot \vec{r}_0^{\mathrm{FI}}}.
\end{equation}
$D_{\vec{k}\vec{q}}^{\mathrm{FI}}$ quantifies the degree of overlap between magnonic and photonic modes. An analogous quantity appears in the cavity--SC coupling in Sec.~\ref{subsubsec:SC}, so we define it via the general expression
\begin{align}
    D_{\vec{l}_M \vec{q}}^M \equiv{}& \frac{e^{i(\vec{l}_M - \vec{q}) \cdot \vec{r}_0^M}}{N_M } \sum_{i \in M} e^{-i(\vec{l}_M - \vec{q}) \cdot \vec{r}_i} \nonumber\\*
    &\times \left\{\begin{array}{l r}
        -\cos\frac{\pi z_i}{L_z}, & M = \mathrm{FM}\\
        \sin\frac{\pi z_i}{L_z}, & M = \mathrm{SC}
        \end{array}\right\} \nonumber\\*
    \approx{}& \delta_{l_{M,z} 0} \prod_d \text{sinc}\left[\pi N_d^M\left( \frac{m_d^M}{N_d^M} - \frac{\ell_d a_M}{L_d} \right)\right]. \label{eq:DsumFI}
\end{align}
Here $M = \{\mathrm{FI}, \mathrm{SC}\}$ is a material index, $\vec{l}_M$ represents either a magnon or an SC quasiparticle momentum, $\vec{r}_0^M$ is the center position of lattice $M$ relative to the origin, and the photon momentum numbers $\ell_d = \ell_x, \ell_y$ were defined under Eq.~\eqref{eq:Qvec}. The latter, along with other SC quantities, are defined in Sec.~\ref{subsubsec:SC}. The sum over $i$ is taken over either FI or SC lattice points, as indicated by $M$, and the last equality holds for $N_d^M \gg 1$.

$D_{\vec{k} \vec{q}}^{\mathrm{FI}}$ reduces to a Kronecker delta $\delta_{\vec{k} \vec{q}}$ only when $L_{\p} = l_{\p} = a_{\mathrm{FI}} N_{\p}^{\mathrm{FI}}$, \ie when the FI and the cavity share in-plane dimensions~\footnote{More precisely, $D_{\vec{k} \vec{q}}^{\mathrm{FI}}$ equals an infinite sum of Kronecker delta functions when the FI and the cavity share in-plane dimensions: one for each $\vec{q}$ that is equivalent to $\vec{k}$ up to an FI Brillouin zone. We are anyhow only concerned with the first Brillouin zone, since the interaction strengths decrease rapidly with increasing $|\vec{q}|$ due to factors $\omega_{\vec{q}}^{-1}$ entering the coupling strengths.}. At the other end of the scale, when the FI becomes infinitely small, $D_{\vec{k} \vec{q}}^{\mathrm{FI}}$ reduces to $\delta_{\vec{k} \vec{0}}$, implying all cavity modes couple exclusively to the uniform magnon mode, which is often assumed in cavity implementations~\cite{Janssonn2020, Johansen2018, Tabuchi2016, Huebl2013}. We assume this uniform coupling only in the $z$ direction, hence the factor $\delta_{l_{M,z} 0}$ in Eq.~\eqref{eq:DsumFI} (thus $k_z = 0$); the condition is that $\pi d_M / 2 L_z \ll 1$, with $d_M$ the thickness of film $M$~\footnote{More generally, up to an overall sign, the condition is that $\pi d_M \ell_z / 2 L_z \ll 1$ with $\ell_z$ odd; even $\ell_z$ cavity modes do not have in-plane electric field components at the location of the SC, leaving them uninteresting for our purposes. This condition is seen to require increasingly thin films with higher $\ell_z$. However, higher $\ell_z$ cavity modes enter interactions at increasingly great energy costs, leaving $\ell_z = 1$ modes the predominant modes entering our interactions owing to the geometry and configuration of our set-up.}.

\subsubsection{Superconductor} \label{subsubsec:SC}

The SC Hamiltonian is
\begin{align}
    \mathcal{H}_{\mathrm{SC}} & = \mathcal{H}_{\mathrm{sing}} + \mathcal{H}_{\mathrm{BCS}} + \mathcal{H}_{\mathrm{para}}, \label{eq:HSCinitial}
\end{align}
with
\begin{subequations}
\begin{align}
    \mathcal{H}_{\mathrm{sing}} & \equiv \sum_{\vec{p}}  \xi_{\vec{p}}c_{\vec{p}\sigma}^\dagger c_{\vec{p}\sigma'}, \label{eq:Hsinginitial} \\
    \mathcal{H}_{\mathrm{BCS}} & \equiv - \sum_{\vec{p}} \left( \Delta_{\vec{p}} c_{\vec{p} + \vec{P}, \uparrow}^\dagger c_{-\vec{p} + \vec{P}, \downarrow}^\dagger + \Delta_{\vec{p}}^* c_{-\vec{p} + \vec{P}, \downarrow} c_{\vec{p} + \vec{P}, \uparrow} \right), \label{eq:HBCSInit} \\
    \mathcal{H}_{\mathrm{para}} & \equiv \sum_{\p}\sum_{j} j_{\p}(\vec{r}_j) A_{\p}\left(\frac{\vec{r}_{j + I_{\p}} + \vec{r}_{j}}{2}\right), \label{eq:HparaInit}
\end{align}
\end{subequations}
$\mathcal{H}_{\mathrm{sing}}$ is the single-particle energy, where $\xi_{\vec{p}}$ is the lattice-dependent electron dispersion, and $c_{\vec{p}\sigma}$ and $c_{\vec{p}\sigma}^\dagger$ are fermionic operators for an electron of lattice momentum $\vec{p}$ and spin $\sigma$. The momenta are discretized as
\begin{equation} \label{eq:MomSC}
    \vec{p} \equiv (2\pi m_x^{\mathrm{SC}}/l_x^{\mathrm{SC}}, 2\pi m_y^{\mathrm{SC}}/l_y^{\mathrm{SC}}, 2\pi m_z^{\mathrm{SC}}/l_z^{\mathrm{SC}}) \equiv (p_x, p_y, p_z),
\end{equation}
where $m_{\p}^{\mathrm{SC}}$ and $m_z^{\mathrm{SC}}$ are defined analogously to $m_{\p}^{\mathrm{FI}}$ (see below Eq.~\eqref{eq:MomFI}), covering the 1BZ of the SC with $N_{\p}^{\mathrm{SC}}$ ($N_{z}^{\mathrm{SC}}$) the number of SC lattice points in direction $\p$ ($z$).

$\mathcal{H}_{\mathrm{BCS}}$ is the BCS pairing term, with $\Delta_{\vec{p}}$ the pairing potential. The leading order effect of applying an in-plane DC across the SC is to shift the center of the SC pairing potential from $\vec{p} = \vec{0}$ to $\vec{p} = \vec{P}$, where $2\vec{P}$ is the generally finite center-of-mass momentum of the Cooper pairs~\cite{Takashima2017,Johnsen2021,Takasan}. The maximum value of $\vec{P}$ is limited by the critical current of the superconductor.

$ \mathcal{H}_{\mathrm{para}} $ is the paramagnetic coupling. $j_{\p}(\vec{r}_j)$ is the $\p$ component of the discretized electric current operator at lattice site $j$ with the position vector $\vec{r}_j$, and is defined as \cite{Schlawin2019}
\begin{equation} \label{eq:SCelectriccurrent}
    j_{\p}(\vec{r}_j) \equiv \frac{ia_{\mathrm{SC}}et}{\hbar} \sum_\sigma ( c_{j+I_{\p},\sigma}^\dagger c_{j\sigma} - c_{j\sigma}^\dagger c_{j+I_{\p},\sigma} ).
\end{equation}
The $z$ component $j_{z}$ does not contribute to our Hamiltonian because the cavity gauge field is in-plane at $z \approx L_z/2$. Above, $a_{\mathrm{SC}}$ is the lattice constant, $e$ is the electric charge, $t$ is the lattice hopping parameter, and $c_{j\sigma}$ and $c_{j\sigma}^\dagger$ are real-space fermionic operators for electrons with spin $\sigma$ at lattice site $j$. They relate to $c_{\vec{p}\sigma}$ and $c_{\vec{p}\sigma}^\dagger$ via
\begin{equation} \label{eq:electronsfourier}
\begin{split}
    c_{j\sigma} = \frac{1}{\sqrt{N_{\mathrm{SC}}}} \sum_{\vec{p}} c_{\vec{p}\sigma} e^{i\vec{p}\cdot \vec{r}_j},
\end{split}
\end{equation}
with $N_{\mathrm{SC}}$ the total number of SC lattice points. Furthermore, $I_{\p}$ represents a unit step in the $\p$ direction with respect to lattice labels. For instance, if $j = (1,1)$, then $j + I_x = (1+1,1) = (2,1)$.

Inserting Eqs.~\eqref{eq:Acavgauged}, \eqref{eq:SCelectriccurrent} and \eqref{eq:electronsfourier} into Eq.~\eqref{eq:HparaInit} yields
\begin{equation} \label{eq:Hparafull}
\begin{split}
    \mathcal{H}_{\mathrm{para}} = \sum_{\vec{p} \vec{p}^\prime \sigma} \sum_{\vec{q} \varsigma} g_{\varsigma}^{\vec{q} \vec{p} \vec{p}^\prime} ( a_{\vec{q}  \varsigma} + a_{-\vec{q}  \varsigma}^\dagger ) c_{\vec{p} \sigma}^\dagger c_{\vec{p}^\prime \sigma}.
\end{split} 
\end{equation}
Here, we have introduced the coupling strength
\begin{align} \label{eq:SCcavcouplingconstant}
    g_{\varsigma}^{\vec{q} \vec{p} \vec{p}^\prime} \equiv & -\frac{a_{\mathrm{SC}}et}{\hbar} \sqrt{\frac{\hbar}{\epsilon \omega_{\vec{q}}V}} D_{\vec{p} - \vec{p}^\prime, \vec{q}}^{\mathrm{SC}} e^{i\vec{q}\cdot \vec{r}_0^{\mathrm{SC}}} \nonumber\\
    & \cdot \sum_{\p} \left( e^{-i(\vec{p} - \vec{q}/2 )\cdot \bm{\delta}_{\p}} - e^{i(\vec{p}^\prime + \vec{q}/2) \cdot \bm{\delta}_{\p}} \right) O^{\vec{q}}_{ \varsigma \p},
\end{align}
where $\bm{\delta}_{\p} \equiv a_{\mathrm{SC}} \hat{e}_d$ are in-plane primitive lattice vectors. $D_{\vec{p}-\vec{p}^\prime, \vec{q}}^{\mathrm{SC}}$ is defined in Eq.~\eqref{eq:DsumFI}, quantifying the degree of overlap between two electron modes and a photon mode. It reduces to $\delta_{\vec{p} - \vec{p}^\prime, \vec{q}}$ only when the cavity and the SC share in-plane dimensions, as is the case in Ref.~\cite{Schlawin2019}.

As we move onto the imaginary-time (Matsubara) path integral formalism in the next sections, it becomes convenient to eliminate creation--creation and annihilation--annihilation fermionic operator products. To this end, we absorb the BCS term~\eqref{eq:HBCSInit} into the diagonal term~\eqref{eq:Hsinginitial} by a straight-forward diagonalization:
\begin{align}
    & \mathcal{H}_{\mathrm{sing}} + \mathcal{H}_{\mathrm{BCS}} \nonumber\\*
    & = \sum_{\vec{p}}
    \begin{pmatrix}
    c_{\vec{p} + \vec{P}, \uparrow} \\
    c_{-\vec{p} + \vec{P}, \downarrow}^\dagger
    \end{pmatrix}^\dagger
    \begin{pmatrix}
    \xi_{\vec{p} + \vec{P}} & -\Delta_{\vec{p}} \\
    -\Delta_{\vec{p}}^* & -\xi_{-\vec{p} + \vec{P}}
    \end{pmatrix}
    \begin{pmatrix}
    c_{\vec{p} + \vec{P}, \uparrow} \\
    c_{-\vec{p} + \vec{P}, \downarrow}^\dagger
    \end{pmatrix} \nonumber\\*
    & = \sum_{\vec{p}}
    \begin{pmatrix}
    \gamma_{\vec{p} 0} \\
    \gamma_{\vec{p} 1}
    \end{pmatrix}^\dagger
    \begin{pmatrix}
    E_{\vec{p} 0} & 0 \\
    0 & E_{\vec{p} 1}
    \end{pmatrix}
    \begin{pmatrix}
    \gamma_{\vec{p} 0} \\
    \gamma_{\vec{p} 1}
    \end{pmatrix}. \label{eq:HSCDCNoPara}
\end{align}
Here we introduced the Bogoliubov (SC) quasiparticle basis $\{ \gamma_{\vec{p} m} , \gamma_{\vec{p} m}^\dagger \}$, with $m = 0,1$ and dispersion relations
\begin{align}
    E_{\vec{p} m} = & \frac{1}{2} \bigg[ \xi_{\vec{p} + \vec{P}} - \xi_{-\vec{p} + \vec{P}} \nonumber\\*
    & + (-1)^m \sqrt{\left( \xi_{\vec{p} + \vec{P}} + \xi_{-\vec{p} + \vec{P}} \right)^2 + 4 |\Delta_{\vec{p}}|^2 } \bigg]. \label{eq:BQDispersionDC}
\end{align}
The elements $u_{\vec{p}}$ and $v_{\vec{p}}$ of the basis transformation matrix are defined through~\cite{Tinkham}
\begin{equation} \label{eq:BogoliubovQuasiparticlesDC}
    c_{\vec{p} + \vec{P}, \uparrow} \equiv u^*_{\vec{p}} \gamma_{\vec{p}0} + v_{\vec{p}} \gamma_{\vec{p}1}, \quad c_{-\vec{p} + \vec{P}, \downarrow}^\dagger \equiv -v^*_{\vec{p}} \gamma_{\vec{p}0} + u_{\vec{p}} \gamma_{\vec{p}1}.
\end{equation}
Inserting the above into Eq.~\eqref{eq:HSCDCNoPara}, one finds the relations
\begin{subequations}
\begin{align} 
    \frac{\Delta_{\vec{p}}^* v_{\vec{p}}}{u_{\vec{p}}} ={}& \frac{1}{2} \left[ \left( E_{\vec{p} 0} - E_{\vec{p} 1} \right) - \left( \xi_{\vec{p} + \vec{P}} + \xi_{-\vec{p} + \vec{P}} \right) \right],\label{eq:BCSdiagonalizationconditionDC} \\
    |v_{\vec{p}}|^2 ={}& 1 - |u_{\vec{p}}|^2 = \frac{1}{2}\left( 1 - \frac{\xi_{\vec{p} + \vec{P}} + \xi_{-\vec{p} + \vec{P}}}{E_{\vec{p} 0} - E_{\vec{p} 1}} \right),
\end{align}
\end{subequations}
which determine $u_{\vec{p}}$ and $v_{\vec{p}}$. Recasting $\mathcal{H}_{\mathrm{para}}$ in terms of this basis yields
\begin{equation}
\begin{split}
    \mathcal{H}_{\mathrm{para}} = & \sum_{\vec{p} \vec{p}^\prime} \sum_{\vec{q} \varsigma} \sum_{mm^\prime} g_{\varsigma m m^\prime}^{\vec{q} \vec{p} \vec{p}^\prime} ( a_{\vec{q}  \varsigma} + a_{-\vec{q}  \varsigma}^\dagger ) \gamma_{\vec{p}m}^\dagger \gamma_{\vec{p}^\prime m^\prime},
\end{split} 
\end{equation}
where the coupling strength is now
\begin{widetext}
\begin{equation} \label{eq:SCcavcouplingQuasiparticlesDC}
\begin{split}
    g_{\varsigma m m^\prime}^{\vec{q} \vec{p} \vec{p}^\prime} \equiv &
    \begin{pmatrix}
    g_{\varsigma}^{\vec{q}, \vec{p} + \vec{P}, \vec{p}^\prime + \vec{P}} u_{\vec{p}} u_{\vec{p}^\prime}^* + g_{\varsigma}^{\vec{q}, \vec{p} - \vec{P}, \vec{p}^\prime - \vec{P}} v_{\vec{p}} v_{\vec{p}^\prime}^* & g_{\varsigma}^{\vec{q}, \vec{p} + \vec{P}, \vec{p}^\prime + \vec{P}} u_{\vec{p}} v_{\vec{p}^\prime} - g_{\varsigma}^{\vec{q}, \vec{p} - \vec{P}, \vec{p}^\prime - \vec{P}}v_{\vec{p}} u_{\vec{p}^\prime} \\
    -g_{\varsigma}^{\vec{q}, \vec{p} - \vec{P}, \vec{p}^\prime - \vec{P}} u_{\vec{p}}^* v_{\vec{p}^\prime}^* + g_{\varsigma}^{\vec{q}, \vec{p} + \vec{P}, \vec{p}^\prime + \vec{P}} v_{\vec{p}}^* u_{\vec{p}^\prime}^* & g_{\varsigma}^{\vec{q}, \vec{p} - \vec{P}, \vec{p}^\prime - \vec{P}} u_{\vec{p}}^* u_{\vec{p}^\prime} + g_{\varsigma}^{\vec{q}, \vec{p} + \vec{P}, \vec{p}^\prime + \vec{P}} v_{\vec{p}}^* v_{\vec{p}^\prime}
    \end{pmatrix}_{m m^\prime}.
\end{split}
\end{equation}
\end{widetext}

This concludes the derivation of the terms entering the system Hamiltonian in terms of the various (quasi)particle bases. We now turn our focus to the construction of an effective FI theory.

\subsection{Imaginary time path integral formalism} \label{sec:Matsubara}

We now seek to extract the influence of the SC on the FI, in particular the anisotropy field induced across the FI. Diagonalizing the Hamiltonian directly, as was done in Eq.~\eqref{eq:HSCDCNoPara}, would in this case be very challenging, as it couples many more modes, and furthermore contains trilinear operator products. Since the external drives ($\vec{B}_{\mathrm{ext}}$ and the DC) only give rise to equilibrium phenomena in our system, the Matsubara path integral formalism of evaluating thermal correlation functions is valid~\cite{Altland}. This translates the evaluation into a path integral problem, which is very convenient for our purposes. The path integral approach facilitates aggregation of the influences of specific subsystems into effective actions, without explicit diagonalization. On this note, for comparison, \citet{Cottet2020} analyze a scenario in which the non-equilibrium Keldysh path integral formalism is used to analyze the net influence of a QED circuit on a cavity.

The starting point is the imaginary time action
\begin{align} \label{eq:action}
    S \equiv{}& S^{\mathrm{FI}}_0 + S^{\mathrm{cav}}_0 + S^{\mathrm{SC}}_0 + S^{\mathrm{FI-cav}}_{\mathrm{int}} + S^{\mathrm{cav-SC}}_{\mathrm{int}} \nonumber\\*
    ={}& \int\mathrm{d}\tau \bigg[\sum_{\vk}\eta_{\vk}^\dagger \hbar\partial_\tau \eta_{\vk} + \sum_{\vq\varsigma}a^\dagger_{\vq \varsigma} \hbar\partial_\tau a_{\vq\varsigma} \nonumber\\*
    &+ \sum_{\vp m}\gamma_{\vp m}^\dagger \hbar\partial_\tau \gamma_{\vp m} + \mathcal{H}\bigg].
\end{align}
$\tau$ is a temperature parameter treated as imaginary time, which relates to the thermal equilibrium density matrix $\exp (-\beta \mathcal{H}/\hbar)$, with $\beta \equiv \hbar/k_{\mathrm{B}} T$ the inverse temperature $T$ in units of time, and $\mathcal{H}$ the system Hamiltonian. The dependence of the field operators on temperature ($\tau$) is implied. In formulating the path integral, the magnon, photon and Bogoliubov quasiparticle operators have been replaced by eigenvalues of the respective coherent states \cite{Altland}; \ie the bosonic operators have been replaced by complex numbers, and the fermionic operators by Gra{\ss}mann numbers. The magnons, photons and Bogoliubov quasiparticles are furthermore taken to be functions of $\tau$ \cite{Altland}. The integral over $\tau$ is taken over the interval $( 0, \beta ]$. Note that we assume the gap to be fixed to the bulk mean field value, and therefore do not include a gap action or integration in the partition function.

We now replace the integral over $\tau$ by an infinite sum over discrete frequencies by a Fourier transform of the magnon, photon and Bogoliubov quasiparticle operators with respect to $\tau$. The conjugate Fourier parameters are Matsubara frequencies:
\begin{equation}
    \Omega_n = \frac{2n\pi}{\beta}
\end{equation}
for bosons, and
\begin{equation}
    \omega_n = \frac{(2n + 1)\pi}{\beta}
\end{equation}
for fermions, with $n \in \mathbb{Z}$. The transforms read
\begin{subequations}
\begin{align}
    \eta_{\vec{k}} & = \frac{1}{\sqrt{\beta}} \sum_{\Omega_m} \eta_{-\Omega_m, \vec{k}} e^{-i\Omega_m \tau}, \label{eq:Matsubaraetak} \\
    a_{\vec{q} \varsigma} & = \frac{1}{\sqrt{\beta}} \sum_{\Omega_n} a_{-\Omega_n, \vec{q} \varsigma} e^{-i\Omega_n \tau}, \\
    \gamma_{\vec{p} m} & = \frac{1}{\sqrt{\beta}} \sum_{\omega_n} \gamma_{- \omega_n, \vec{p} m} e^{-i\omega_n \tau}. \label{eq:Matsubaragammapm}
\end{align}
\end{subequations}
To avoid clutter, we introduce the 4-vectors
\begin{subequations}
\begin{align}
    k & \equiv (-\Omega_m, \vec{k}), \\
    q & \equiv (-\Omega_n, \vec{q}), \\
    p & \equiv (-\omega_n, \vec{p}),
\end{align}
\end{subequations}
and the generally complex energies
\begin{subequations}
\begin{align}
    \hbar\lambda_k & \equiv -i\hbar\Omega_m + \hbar\lambda_{\vec{k}}, \\
    \hbar\omega_q & \equiv -i\hbar\Omega_n + \hbar\omega_{\vec{q}}, \label{eq:omegaqPathInt} \\
    E_{pm} & \equiv -i\hbar\omega_n + E_{\vec{p}m}. \label{eq:EpmPathInt}
\end{align}
\end{subequations}
The actions in~\eqref{eq:action} then become
\begin{subequations}
\begin{align}
    S^{\mathrm{FI}}_0 & = \sum_{k} \hbar \lambda_k \eta_{k}^\dagger \eta_{k}, \label{eq:actionsBareFI} \\
    S^{\mathrm{cav}}_0 & = \sum_{q \varsigma} \hbar \omega_q a_{q \varsigma}^\dagger a_{q \varsigma}, \\
    S^{\mathrm{SC}}_0 & = \sum_{p m} E_{p m} \gamma_{p m}^\dagger \gamma_{p m},  \label{eq:actionsBareSC} \\
    S^{\mathrm{FI-cav}}_{\mathrm{int}} & = \sum_{k \p} \sum_{q \varsigma} g_{\p \varsigma}^{k q} ( \nu_{\p} \eta_{-k} + \nu_{\p}^* \eta_{k}^\dagger) ( a_{q \varsigma} + a_{-q \varsigma}^\dagger ), \label{eq:actionsIntFIcav} \\
    S^{\mathrm{cav}-\mathrm{SC}}_{\mathrm{int}} & = \frac{1}{\sqrt{\beta}}\sum_{q \varsigma} \sum_{p m} \sum_{p^\prime m^\prime} g_{\varsigma m m^\prime}^{q p p^\prime} ( a_{q \varsigma} + a_{-q \varsigma}^\dagger ) \gamma_{p m}^\dagger \gamma_{p^\prime m^\prime}, \label{eq:actionsIntcavSC}
\end{align}
\end{subequations}
where we introduced the coupling functions
\begin{align}
    g_{\p \varsigma}^{k q} & \equiv g_{\p}^{\vec{k} \vec{q}} \delta_{\varsigma 1} \delta_{\Omega_m, \Omega_n}, \label{eq:actiongFI} \\
    g_{\varsigma m m^\prime}^{q p p^\prime} & \equiv g_{\varsigma m m^\prime}^{\vec{q} \vec{p} \vec{p}^\prime} \delta_{\omega_{n^\prime}, \omega_n - \Omega_n}. \label{eq:actiongSC}
\end{align}
We additionally introduced a redundant Kronecker delta function $\delta_{\varsigma 1}$ to the coupling \eqref{eq:actiongFI}, which will facilitate the gathering of interaction terms in Eq.~\eqref{eq:SintcavGathered}. We will use the notation $g^\eta$ and $g^\gamma$ for the magnitudes of the FI--cavity and cavity--SC coupling, respectively.

We are now equipped to construct effective actions by integrating out the photonic and fermionic degrees of freedom, to which end we will consider the imaginary-time partition function~\cite{Altland, Kachelriess}
\begin{align}
    Z & \equiv \langle \mathrm{vac}, t=\infty  \lvert \mathrm{vac}, t=-\infty \rangle \nonumber\\*
    & = \int \mathcal{D}[\eta, \eta^\dagger] \int \mathcal{D}[a, a^\dagger] \int \mathcal{D}[\gamma, \gamma^\dagger] e^{-S/\hbar}, \label{eq:partitionFunc}
\end{align}
where e.g.
\begin{equation}
    \int \mathcal{D}[\gamma, \gamma^\dagger] \equiv \prod_{p m} \int \mathcal{D}[\gamma_{p m}, \gamma_{p m}^\dagger]
\end{equation}
is to be understood as the path integrals over every Bogoliubov quasiparticle mode.

\subsection{Integrating out the cavity photons}
The order in which we integrate out the cavity and the SC is inconsequential. We will begin with the cavity, which can be integrated out exactly. We show that interchanging the order of integrations leads to identical results in \cref{app:SCfirst}.

\begin{figure}
    \includegraphics[width=\columnwidth]{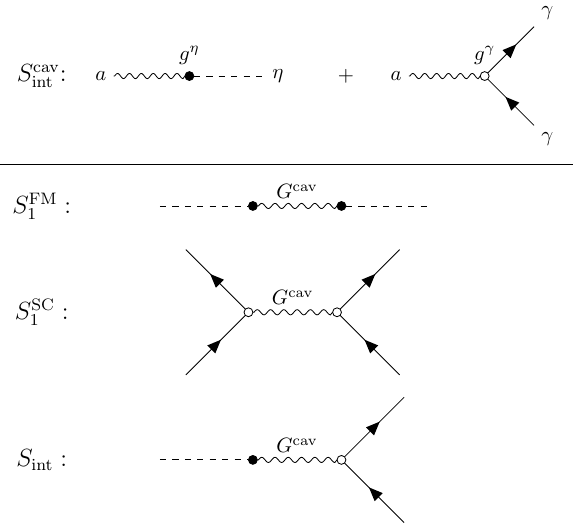}
    \caption{Feynman diagrams \cite{Ellis2017} of the bare cavity coupling to the FI and SC, and the resulting terms in the FI and SC effective actions after integrating out the cavity photons, where $G^\mathrm{cav}$ is the photon propagator.}
    \label{fig:interaction_diagrams}
\end{figure}

We gather the interactions between the cavity and FI and SC,
\begin{align} \label{eq:SintcavGathered}
    S_\mathrm{int}^\mathrm{cav} = \sum_{q,\varsigma} [J_{q\varsigma}a_{q\varsigma} + J_{-q\varsigma}a_{-q\varsigma}^\dagger],
\end{align}
where we have defined
\begin{align}
    J_{q\varsigma} ={}& \sum_{ks} g_{\p\varsigma}^{kq}( \nu_{\p} \eta_{-k} + \nu_{\p}^* \eta_{k}^\dagger) \nonumber\\*
    &+ \frac{1}{\sqrt{\beta}} \sum_{pp'}\sum_{mm'} g_{\varsigma mm'}^{qpp'} \gamma_{pm}^\dagger \gamma_{p'm'}.
\end{align}
These interaction terms are illustrated by the diagrams in the top panel of \cref{fig:interaction_diagrams}. Integrating out the cavity modes \cite{Altland}, we therefore get
the effective action
\begin{align}
    S_\mathrm{eff} = -\sum_{q\varsigma} \frac{J_{q\varsigma}J_{-q\varsigma}}{\hbar\omega_q}.
\end{align}
Inserting the expression for $J_{q\varsigma}$ we get three different terms, $S_\mathrm{eff} = S^\mathrm{FI}_1 + S_1^\mathrm{SC} + S_\mathrm{int}$, shown diagrammatically in the bottom panel of \cref{fig:interaction_diagrams}.  The first term,
\begin{align}
    S_1^\mathrm{FI} ={}& -\sum_{qkk'}\sum_{\varsigma \p\p'}\frac{g_{\p\varsigma}^{kq}g_{\p'\varsigma}^{k'-q}}{\hbar\omega_q} \nonumber\\*
    &\times ( \nu_{\p} \eta_{-k} + \nu_{\p}^* \eta_{k}^\dagger)(\nu_{\p'}\eta_{-k'} + \nu_{\p'}^*\eta_{k'}^\dagger),
\end{align}
is a renormalization of the magnon theory due to interactions with the cavity, resulting in a non-diagonal theory. The second term,
\begin{align}
    S_1^\mathrm{SC} ={}& - \frac{1}{\beta}\sum_{\substack{qpp'\\oo'}}\sum_{\substack{\varsigma mm'\\nn'}} \frac{g_{\varsigma mm'}^{qpp'} g_{\varsigma nn'}^{-qoo'}}{\hbar\omega_q} \gamma_{pm}^\dagger \gamma_{p'm'}\gamma_{on}^\dagger \gamma_{o'n'},
\end{align}
is an interaction term coupling four quasiparticles, similar to the term found in Ref.~\cite{Schlawin2019} for a normal metal coupled to a cavity, leading to superconducting correlations. Note that unlike the pairing term found in Ref.~\cite{Schlawin2019} via the Schrieffer--Wolff transformation, the term above is not limited to an off-resonant regime. In principle it could also lead to renormalization of the quasiparticle spectrum and lifetime. Since we are here concerned with the effects of the cavity and SC on the FI, we will neglect this term as it only leads to higher order corrections.

Finally, we have the cavity-mediated magnon-quasiparticle coupling,
\begin{align}
    S_\mathrm{int} ={}& -\frac{1}{\sqrt{\beta}}\sum_{kpp'}\sum_{\p mm'}V_{\p mm'}^{kpp'} ( \nu_{\p} \eta_{-k} + \nu_{\p}^* \eta_{k}^\dagger)\gamma_{pm}^\dagger \gamma_{p'm'},
\end{align}
where we have defined the effective FI-SC interaction
\begin{align} \label{eq:Vkppdmm}
    V_{\p mm'}^{kpp'} = \sum_{q\varsigma} g^{kq}_{\p\varsigma}g_{\varsigma mm'}^{-qpp'}\l[\frac1{\hbar \omega_q} + \frac1{\hbar\omega_{-q}}\r].
\end{align}
This term is generally nonzero, and we therefore see that the cavity photons lead to a coupling between the FI and SC, potentially over macroscopic distances. This means that the FI and SC will have a mutual influence on each other, possibly leading to experimentally observable changes in the two materials. We therefore integrate out the Bogoliubov quasiparticles and calculate the effective FI theory below. We reiterate that the interaction is exact at this point, not a result of a perturbative expansion.

\subsection{Integrating out the SC quasiparticles --- effective FI theory} \label{sec:IntOutSC}
The full effective SC action comprises the sum $S_0^\mathrm{SC} + S_1^\mathrm{SC} + S_\mathrm{int}$. The second term is second order in $g^\gamma$, but does not contain FI operators, and will therefore only have an indirect effect on the effective FI action. In a perturbation expansion of the effective FI action, the term $S_1^\mathrm{SC}$ will therefore contribute higher order correction terms compared to $S_\mathrm{int}$. We therefore neglect this term in the following, leading to the SC action
\begin{align}
    S^\mathrm{SC} \approx -\sum_{pp'}\sum_{mm'}\gamma_{pm}^\dagger (G^{-1})_{mm'}^{pp'}\gamma_{p'm'},
\end{align}
where we have defined $G^{-1} = G_0^{-1} + \Sigma$, with
\begin{align}
    (G_0^{-1})_{mm'}^{pp'} ={}& -E_{pm}\delta_{pp'}\delta_{mm'}, \\*
    \Sigma_{mm'}^{pp'} ={}&  \frac{1}{\sqrt{\beta}}\sum_{k\p}V_{\p mm'}^{kpp'}( \nu_{\p} \eta_{-k} + \nu_{\p}^* \eta_{k}^\dagger).
\end{align}

Integrating out the SC quasiparticles results in the effective FI action \cite{Altland}
\begin{align}
    S^\mathrm{FI} = S_0^\mathrm{FI} + S_1^\mathrm{FI} - \hbar\operatorname{Tr} \ln (-\beta G^{-1}/\hbar).
\end{align}
The Green's function matrix $G^{-1}$ contains magnon fields, and will be treated perturbatively in order to draw out the lowest order terms in the effective FI theory. We expand the logarithm to second order in the FI--SC interaction,
\begin{align}
    \ln \l(-\frac{\beta G^{-1}}{\hbar}\r) \approx \ln\l(-\frac{\beta G_0^{-1}}{\hbar}\r) + G_0\Sigma - \frac12 G_0\Sigma G_0\Sigma, \label{eq:logexpansion}
\end{align}
where $G_0$ is the inverse of $G_0^{-1}$. This expansion is valid when $|G_0 \Sigma|\ll1$, meaning $|g^\eta g^\gamma/\hbar\omega_q E_{pm}| \ll 1$, where we use shorthand notation for the couplings $g^\eta$ and $g^\gamma$ between cavity photons and $\eta$ and $\gamma$ fields respectively. The first term in \cref{eq:logexpansion} does not contain magnonic fields, and therefore does not contribute to the FI effective action \footnote{For the same reason the term $S_1^\mathrm{SC}$ in the SC action would only contribute when paired with $\Sigma$, leading to terms two orders higher in $g^\gamma$ compared to the terms containing $\Sigma$ only.}. The third term contains bilinear terms in magnonic fields, and gives a correction to the magnon dispersion of order $|[g^\eta g^\gamma/\hbar\omega_q]^2/ E_{pm}|$, a factor of $|(g^\gamma)^2/\hbar\omega_q E_{pm}|$ smaller than the corrections contained in $S_1^\mathrm{FI}$, and will therefore also be neglected. Keeping only the second term, and using the fact that $G_0$ is diagonal in both quasiparticle type $m$ and momentum $p$, we therefore get the effective FI action to leading order,
\begin{align}
    S^\mathrm{FI} ={}& \sum_k \hbar\lambda_k \eta_k^\dagger \eta_k -g\mu_B\sum_{k\p} h_\p^k \cdot \sqrt{\frac{S}{2}} ( \nu_{\p} \eta_{-k} + \nu_{\p}^* \eta_{k}^\dagger) \nonumber\\*
    &+ \sum_{kk'\p\p'} Q_{\p\p'}^{kk'}( \nu_{\p} \eta_{-k} + \nu_{\p}^* \eta_{k}^\dagger) ( \nu_{\p'} \eta_{-k'} + \nu_{\p'}^* \eta_{k'}^\dagger),
    \label{eq:SFIeff}
\end{align}
where we have defined the anisotropy field due to the coupling to the superconductor,
\begin{align} \label{eq:hkd}
    h^k_\p ={}& -\frac{\hbar}{g\mu_B} \sqrt{\frac{2}{S\beta}}\sum_{pm} \frac{V_{\p mm}^{kpp}}{E_{pm}},
\end{align}
and a function
\begin{align} \label{eq:Qkkdd}
     Q_{\p\p'}^{kk'} \equiv{}& - \sum_{q\varsigma} \frac{g_{\p\varsigma}^{kq}g_{\p'\varsigma}^{k'-q}}{\hbar\omega_q}.
\end{align}
describing the cavity-mediated self-interaction in the ferromagnetic insulator.

\section{Results}\label{subsec:result_electric}

The main result of our work is the effective magnon action~\eqref{eq:SFIeff}. The interaction with the cavity and the SC gives rise to linear and bilinear correction terms to the diagonal magnon theory, corresponding to an induced anisotropy field and corrections to the magnon spectra.

To extract a specific quantity, we consider the leading order effect of coupling the FI to the SC via the cavity, namely the linear magnon term. Physically this can be understood as a contribution from an additional magnetic field trying to reorient the FI in a direction other than along the $z$ axis. We can see this explicitly if we Fourier transform the linear magnon term back to real space and imaginary time,
\begin{align}\label{eq:SFIlin}
    S_\mathrm{lin}^\mathrm{FI} ={}& -\frac{g\mu_B}{\hbar}\int d\tau \sum_{\vec{r}_i} \sum_{\p} h_{\p}(\vec{r}_i,\tau) S_{id}(\tau),
\end{align}
where we have used the definition of the in-plane spin components in \cref{eq:Sipmsimplecartesian2}, and defined the real space anisotropy field components due to the interaction with the superconductor
\begin{align} \label{eq:hdhkd}
    h_{\p}(\vec{r}_i,\tau) = \frac{1}{\sqrt{N_\mathrm{FI} \beta}} \sum_k h_{\p}^k e^{ik \cdot r_i}.
\end{align}
Above, we introduced the 4-vector
\begin{equation} \label{eq:3vecr}
    r_i \equiv (\tau, \vec{r}_i).
\end{equation}
In order for the anisotropy field components to be real, we require $h_d^k = (h_d^{-k})^*$. Inserting the expressions for $E_{pm}$ and $V_{dmm}^{kpp}$ from \cref{eq:Vkppdmm,eq:EpmPathInt} into \cref{eq:hkd}, and performing the sum over the Matsubara frequencies \cite{Altland}, we get the following expression for the Fourier transposed anisotropy field components,
\begin{align}
    h_d^k ={}& -\sqrt{N_\mathrm{FI}\beta}\delta_{\Om0}\sum_{\vq,d'}\frac{4\pi a_\mathrm{SC}et}{\hbar\epsilon \omega_\vq^2 VL_z}\frac{q_{\bar{d}} q_{d'}}{|\vec{Q}|^2}\nu_d^2 e^{i\vq\cdot(\vec{r}_0^\mathrm{FI} - \vec{r}_0^\mathrm{SC})}\nonumber\\*
    &\times D_{\vk,\vq}^\mathrm{FI}D_{\vec{0},-\vq}^\mathrm{SC} e^{-iq_{d'}a_\mathrm{SC}/2} \Pi_{\vec{P}d'}, \label{eq:hdkSimplified}
\end{align}
where the dependence on the supercurrent comes in through the factor
\begin{align}
    \Pi_{\vec{P}d} ={}& \sum_{\vp}\big\{\sin[(p_d + P_d)a_\mathrm{SC}]|u_\vp|^2 \nonumber\\*
    &+ \sin[(p_d - P_d)a_\mathrm{SC}] |v_\vp|^2\big\}\tanh\frac{\beta E_{\vp 0}}{2\hbar}. \label{eq:PiPd}
\end{align}
Notice that the field is finite only for zero Matsubara frequency, meaning that it is time-independent (magnetostatic). It is possible to show that $h_d^k = (h_d^{-k})^*$ by letting $\vq \to -\vq$ in the sum in \cref{eq:hdkSimplified}, and using $D_{\vk,\vq}^\mathrm{FI} = (D_{-\vk,-\vq}^\mathrm{FI})^*$, $D_{0,-\vq}^\mathrm{SC} = (D_{0,\vq}^\mathrm{SC})^*$ from the definition in \cref{eq:DsumFI}. Observe that in the case of no DC (\ie $\vec{P} = \vec{0}$), the summand in \cref{eq:PiPd} is odd in $\vec{p}$, and the sum therefore zero, i.e., $\Pi_{\vec{P}d} = 0$ if $P_d = 0$. Hence there is no anisotropy field induced across the FI in the absence of a supercurrent. This stresses the necessity of breaking the inversion symmetry of the SC in order to induce an influence on the FI.

\subsection{Special case: small FM} \label{sec:special_case}

\begin{figure}[h!btp]
    \centering
    \includegraphics[width=1\linewidth]{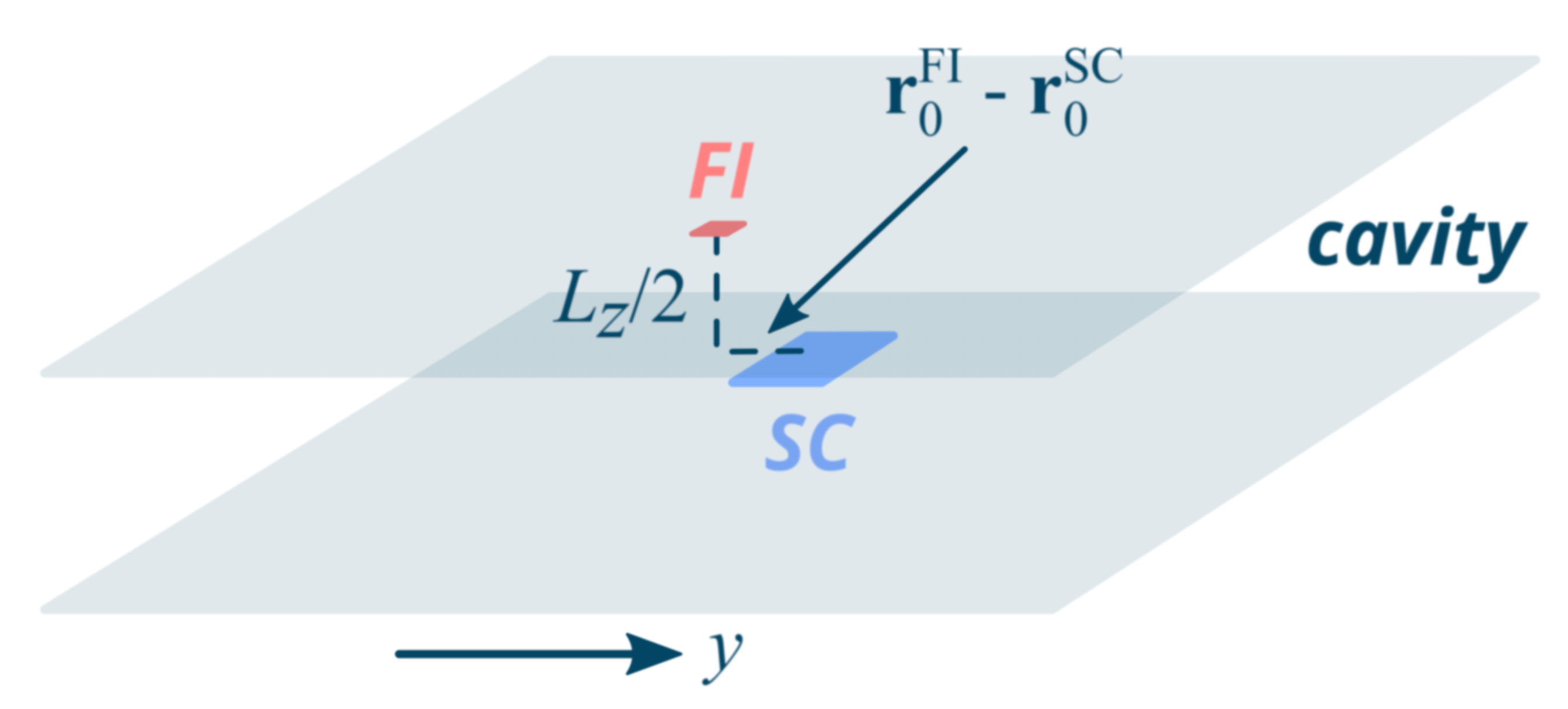}
    \caption{Illustration of the set-up used in the example given in Sec.~\ref{sec:special_case}. A small, square FI and SC are placed spaced apart in the $y$ and $z$ directions inside a comparatively large cavity. Only a small portion of the cavity length in $y$ is utilized as the contributions by the various mediating cavity modes add constructively only over short distances. The FI and SC are nevertheless separated by hundreds of \si{\micro m}, 2--5 orders larger than typical effectual lengths in proximity systems.}
    \label{fig:set-up_anisotropy}
\end{figure}

The anisotropy field~\eqref{eq:hdhkd} generally gives rise to complicated, local reorientation of the FI spins. However, there are special cases in which it takes on a simple form. In particular, assume the FI to be very small relative to the cavity, \ie $\ell_x l^{\mathrm{FI}}_x, \ell_y l^{\mathrm{FI}}_y \ll L_x, L_y$. In this case, the FI sum~\eqref{eq:DsumFI} becomes highly localized around $\vec{k} = \vec{0}$ for the relevant ranges of $\ell_x$ and $\ell_y$, which are limited by the other factors $D_{\vec{0}\vec{q}}^{\mathrm{SC}}$ and $(\omega_{\vec{q}}|\vec{Q}|)^{-2}$ found in \cref{eq:hdkSimplified}. We may therefore set $\vec{k} = \vec{0}$. For a specified set of material parameters and dimensions, the validity is confirmed numerically. In this case, Eq.~\eqref{eq:hdhkd} thus reduces to
\begin{align} \label{eq:hdhkd_smallFI}
    h_{\p} = \frac{h_{\p}^0}{\sqrt{N_\mathrm{FI} \beta}},
\end{align}
representing a uniform anisotropy field across the FI. In this limit we can simplify the expression for the anisotropy field components,
\begin{align}
    h_d ={}& -\sum_{\vq,d'} \frac{2\pi a_\mathrm{SC}e t}{\hbar \epsilon \omega_\vq^2 V L_z} \nu_d^2 D_{\vec{0},\vq}^\mathrm{FI} D_{\vec{0},-\vq}^\mathrm{SC} \Pi_{\vec{P}d'} \frac{q_{\bar{d}}q_{d'}}{|\vec{Q}|^2} \nonumber\\*
    &\times\l[\cos q_x L_x^\mathrm{sep} \cos q_y L_y^\mathrm{sep} - \sin q_x L_x^\mathrm{sep} \sin q_y L_y^\mathrm{sep}\r], \label{eq:h_smallFI}
\end{align}
where we have assumed $e^{-iq_{d'}a_\mathrm{SC}/2} \approx 1$, which is a good approximation as long as the cavity dimensions far exceed the lattice constant and only low $|\vq|$ contribute to the sum, and used the fact that $D_{\vec{0},\vq}^M$ [\cref{eq:DsumFI}] is an even function in $\vq$. We have also defined the separation length $L_d^\mathrm{sep} = (\vec{r}_0^\mathrm{FI} - \vec{r}_0^\mathrm{SC})\cdot \hat{e}_d$. Assuming a finite separation between the FI and SC only in one direction, the last term in the above equation vanishes, making every remaining factor even in $q_d$, except the product $q_{\bar{d}}q_{d'}$ for $\bar{d} \neq d'$. The sum over $\vq$ therefore picks out terms such that $\bar{d} = d'$. In order to get a finite $h_d$ we must, therefore, have $\Pi_{\vec{P}\bar{d}} \neq 0$, i.e., the supercurrent momentum must be finite in the direction $\bar{d}$. Hence, in the case that the separation between the FI and SC is finite in only one direction, applying a supercurrent in the $x$ direction can only induce an anisotropy field in the $y$ direction, and vice versa.

\begin{figure}[h!btp]
\centering
\begin{tabular}[b]{c}
\subfloat[\label{fig:Lsep1}]{\includegraphics[width =.8\linewidth]{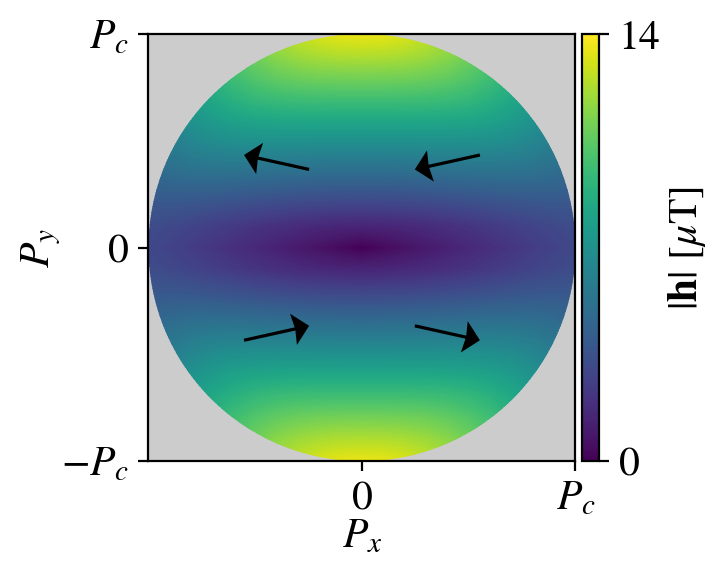}} \\
\subfloat[\label{fig:Lsep2}]{\includegraphics[width =.8\linewidth]{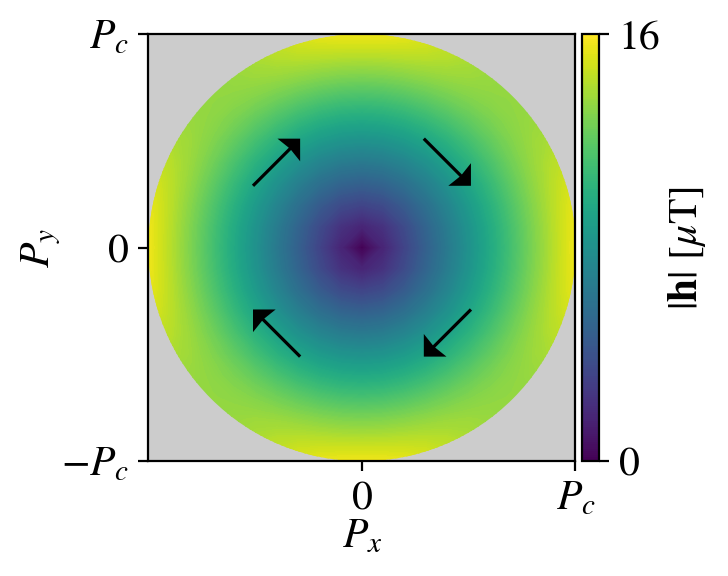}}
\end{tabular}
\caption{The magnitude and direction (arrows) of the effective anisotropy field [\cref{eq:h_smallFI}] at $T = \SI{1}{K}$ as a function of the supercurrent momentum $\vec{P}$, for the simple case of a small FI ($l^{\mathrm{FI}}_x = l^{\mathrm{FI}}_y = \SI{10}{\micro m}$) relative to the cavity ($L_x = L_y = \SI{10}{cm}$, $L_z = \SI{0.1}{mm}$). The SC dimensions are $l^{\mathrm{SC}}_x = l^{\mathrm{SC}}_y = \SI{50}{\micro\meter}$, with a depth of $d_{\mathrm{SC}} = \SI{10}{nm}$. The FI and SC center points are separated by (a) $L_y^{\mathrm{sep}} = \SI{140}{\micro\meter}$ and (b) nothing (placed directly over each other). Observe the change in both the strength and direction of the anisotropy field. The plots were produced using Python with the \texttt{NumPy} and \texttt{Matplotlib} libraries.
}
\label{fig:ax_of_quant}
\end{figure}

We consider the specific case of a small, square FI and SC displaced along $y$ and $z$ (\cref{fig:set-up_anisotropy}). In Fig.~\ref{fig:ax_of_quant} we show numerically how the effective anisotropy field varies with the supercurrent momentum in this special case, using Nb and YIG as material choices for the FI ($l^{\mathrm{FI}}_x = l^{\mathrm{FI}}_y = \SI{10}{\micro m}$) and SC ($l^{\mathrm{SC}}_x = l^{\mathrm{SC}}_y = \SI{50}{\micro\meter}$, $d_{\mathrm{SC}} = \SI{10}{nm}$) films, respectively; see \cref{tab:parameters}. We use Python with the \texttt{NumPy} and \texttt{Matplotlib} libraries for the numerics. We furthermore use the interpolation formula \cite{Gross1986}
\begin{equation} \label{eq:gapInterpolation}
    \Delta = 1.76 k_B T_{c0} \tanh(1.74\sqrt{T_{c0}/T - 1})
\end{equation}
for the superconducting gap, and a simple cubic tight-binding electron dispersion. With the FI and SC center points separated by $\SI{140}{\micro\meter}$ in the $y$ direction (meaning they are separated edge-to-edge by $\SI{115}{\micro\meter}$ in-plane), we find an anisotropy field with a magnitude of $\lesssim \SI{14}{\micro T}$ (Fig.~\ref{fig:Lsep1}). If the constraint on separating the FI and SC in-plane is eased, the maximum magnitude increases to $\SI{16}{\micro T}$ in our specific example (Fig.~\ref{fig:Lsep2}). We discuss the latter case in the concluding remarks.

Two factors determine the inhomogeneous distribution of the responses seen in Fig.~\ref{fig:ax_of_quant}. First, the anisotropy field is nearly linear in the components $P_d$ of the supercurrent momentum, which is seen by expanding the anisotropy field (see~\cref{eq:PiPd}) around $P_d a_{\mathrm{SC}} = 0$ (note that $P_c a_{\mathrm{SC}} \approx 0.001$). This generally makes the response stronger for larger $|\vec{P}|$, which is as expected, since it relies on breaking the $\vec{p}$-inversion symmetry. This dependency is evident in Fig.~\ref{fig:ax_of_quant}. 

Second, the factor $e^{i \vec{q} \cdot (\vec{r}_0^{\mathrm{FI}} - \vec{r}_0^{\mathrm{SC}})}$ renders the anisotropy field very sensitive to the separation of the FI and SC center points in the in-plane directions. This factor expresses that cavity modes associated with a range of different in-plane momenta $\vec{q}$ (i.e., spatial oscillations) with a coherent amplitude at no in-plane separation ($\vec{r}_0^{\mathrm{FI}} - \vec{r}_0^{\mathrm{SC}} = 0$), become increasingly decoherent with increasing separation. Eventually, this decoherence causes states in the SC to contribute oppositely, hence destructively, to the effective anisotropy field. The destructive addition at finite separation is limited by the range of low-$\vec{q}$ cavity modes that contribute to the mediated interaction until the coupling is suppressed by the factor ${D_{\vec{0}\vec{q}}^{\mathrm{FI}} D_{\vec{0} \vec{q}}^{\mathrm{SC}*}}/{\omega_{\vec{q}}^2}\vec{Q}^2$, which in turn is determined by the dimensions of the three subsystems. For sufficiently small separations (determined by the contributing range of $\vq$), this oscillation is mild, and can be used to change the polarity of the anisotropy field without extinguishing the response. This is why the polarity of the response component $h_x$ changes between Figs.~\ref{fig:Lsep1} and \ref{fig:Lsep2}.

It is furthermore clear by inspection of~\cref{eq:PiPd} that the main contributions to the anistotropy field come from states near the Fermi surface. Series-expanding the expression in $\vec{P}$, most terms are seen to cancel due the odd symmetry in $\vec{p}$ that was remarked below~\cref{eq:PiPd}. The strongest asymmetry caused by $\vec{P}$ is seen to originate from the factor $\sin\left[ (p_{\p^\prime} + P_{\p^\prime})a_{\mathrm{SC}}\right] \lvert u_{\vec{p}} \rvert^2 + \sin\left[ (p_{\p^\prime} - P_{\p^\prime})a_{\mathrm{SC}}\right] \lvert v_{\vec{p}} \rvert^2$ in the summand, due to the step-like nature of $\lvert u_{\vec{p}} \rvert^2$ and $\lvert v_{\vec{p}} \rvert^2$ near the Fermi surface. This is as expected, since we consider interactions involving the scattering of SC quasiparticles, hence the low-energy events are concentrated near the Fermi surface.

\begin{table}[h!tbp]
  \begin{threeparttable}
    \caption{Table of numerical parameter values.}
        \begin{ruledtabular}
        \begin{tabular}{ l l l l  }
        \multicolumn{2}{c}{YIG (FI)} & \multicolumn{2}{c}{Nb (SC)} \\
        \hline
        $a_{\mathrm{FI}}$   & $\SI{1.240}{nm}$~\cite{Musa2017}  & $a_{\mathrm{SC}}$ & $\SI{0.330}{nm}$~\cite{Ashcroft} \\
        &  & $T_{c0}$ &  $\SI{6}{K}$~\cite{Gubin2005}\\
        & & $t$ & $\SI{0.35}{eV}$\textsuperscript{a} \\
        & & $P_c$ & $\SI{3.1e7}{\per\meter}$\textsuperscript{b} \\
        & & $E_F$ & $\SI{5.32}{eV}$\textsuperscript{c}~\cite{Ashcroft} \\
        \end{tabular}
        \end{ruledtabular}
        \label{tab:parameters}
    \begin{tablenotes}
      \small
      \item \textsuperscript{a}Based on the tight-binding expression $t = \hbar^2/2ma_{\mathrm{SC}}^2$~\cite{Schlawin2019}, with $m$ the effective electron mass.
      \item \textsuperscript{b}Based on $P_c = j_c m / \hbar e n_s$~\cite{Takasan}, with an estimated critical current $j_c = \SI{4}{MA\per\centi\meter\squared}$~\cite{Ilin2010}, and a superfluid density $n_s = m / \mu_0 e^2 \lambda^2 $~\cite{Tinkham} with a penetration depth $\lambda = \SI{200}{nm}$~\cite{Gubin2005}.
      \item \textsuperscript{c}Fermi energy for Nb. Does not appear explicitly in Eq.~\eqref{eq:h_smallFI}, but is used in the electron dispersion.
    \end{tablenotes}
  \end{threeparttable}
\end{table}

\section{Concluding remarks}\label{sec:conclusion}

In this paper, we have calculated the cavity-mediated coupling between an FI and an SC by exactly integrating out the cavity photons. The main result is the effective FI action~\eqref{eq:SFIeff}, in which linear and bilinear magnon terms appear in addition to the diagonal terms. These respectively correspond to an induced anisotropy field, and corrections to the magnon spectra. In contrast to conventional proximity systems, the cavity-mediation allows for relatively long-distance interactions between the FI and the SC, without destructive effects on order parameters associated with proximity systems, such as pair-breaking magnetic fields. The separation furthermore facilitates subjection of the FI and the SC to separate drives and temperatures. In contrast to common perturbative approaches to cavity-mediated interactions involving the Schrieffer--Wolff transformation~\cite{Schlawin2019, Johansen2018, Johansen2019} or Jaynes--Cummings-like models~\cite{Tabuchi2015, Tabuchi2016, Lambert2016}, the path-integral approach allows for an exact integrating-out of the cavity, without limitations to off-resonant regimes. This carries the additional advantage of allowing for magnon--photon hybridization; that is, we are not theoretically limited to regimes of weak FI--cavity Zeeman coupling. We furthermore take into account that the finite and different FI, cavity and SC dimensions enable interactions between large ranges of particle modes, which is neglected in various preceding works~\cite{Janssonn2020, Johansen2018, Bourhill2016, Tabuchi2016, Huebl2013, Soykal2010, Schlawin2019}, although its importance has been emphasized by both experimentalists \cite{Bourhill2016} and theorists \cite{Soykal2010}.

In an arbitrary practical example, we estimate numerically the effective anisotropy field induced by leading-order interactions across a small YIG film (FI) ($l^{\mathrm{FI}}_x = l^{\mathrm{FI}}_y = \SI{10}{\micro m}$) due to mediated interactions with an Nb film (SC) ($l^{\mathrm{SC}}_x = l^{\mathrm{SC}}_y = \SI{50}{\micro\meter}$, $d_{\mathrm{SC}} = \SI{10}{nm}$). We find it is $\lesssim \SI{14}{\micro T}$, mediated across $\SI{130}{\micro\meter}$ edge-to-edge accounting for both in-plane and out-of-plane separation, inside a $\SI{10}{cm} \times \SI{10}{cm} \times \SI{0.1}{mm}$ cavity (Fig.~\ref{fig:Lsep1}). With out-of-plane coercivities in \si{nm}-thin Bi-doped YIG films reportedly as low as $\SI{300}{\micro T}$~\cite{Lin2020}, this result is expected to yield an experimentally appreciable tilt in the FI spins. The separation is 2--5 orders of magnitude greater than the typical length scales of influence in proximity systems, and facilitates local subjection to different drives and temperatures. The main contributions from the SC originate from a narrow vicinity of the Fermi surface determined by the Cooper pair center-of-mass momentum $2\vec{P}$. The response is very sensitive to the in-plane separation of the FI and SC center points due to the spatial decoherence of the mediating cavity modes over distances, which in turn depends on the dimensions of the FI, cavity and SC. For this reason, the in-plane separation of FI and SC was much smaller than the cavity width.

In \cref{subsec:results_magnetic} we have included the calculation of the anisotropy field when placing the SC at the magnetic antinode at $z=0$. Since the vector potential is purely out of plane in this case, the paramagnetic coupling is zero, and we therefore couple the cavity to the SC via the Zeeman coupling. As shown in the appendix, this results in a much weaker coupling and therefore much smaller anisotropy field. This can be understood by comparing the effective fields the SC couples to in the two cases. The strength of the Zeeman coupling is proportional to $\vq\times \vec{A}$, which for the lowest cavity modes gives a field strength proportional to $|\vec{A}|/L$. However, for the paramagnetic coupling, the effective field is proportional to $\vp \cdot \vec{A}$. In both cases, the main contribution to the anisotropy field originates from a narrow vicinity of the Fermi level, the extent of which is determined by the magnitude of the symmetry-breaking supercurrent (electric antinode) or applied field (magnetic antinode). Thus, we have a paramagnetic coupling proportional to $p_\mathrm{F}|\vec{A}|$, where $p_\mathrm{F}$ is the Fermi momentum. A Fermi energy of $\SI{5.32}{eV}$ gives $p_F \sim \SI{e10}{\per\meter} \gg 1/L$ for cavities with lengths in the \si{mm} to \si{cm} range. Together with the fact that the contributing components of $\vec{A}$ are larger for low $|\vq|$ at the electric antinode compared with the magnetic antinode, the difference in length scales leads to a much larger paramagnetic coupling between cavity and SC compared to the Zeeman coupling, resulting in a much larger effective FI--SC coupling and anisotropy field.

One important constraint in our model that can potentially be eased, is that the FI and the SC cannot overlap in-plane. In this case, we found a stronger response (cf. Fig.~\ref{fig:Lsep2}). This was assumed in order to enable the FI to be subjected to the aligning magnetostatic field $\vec{B}_{\mathrm{ext}}$ without affecting the SC, analogously to the experimental set-up in Refs.~\cite{Tabuchi2015, Tabuchi2016}. Combined with the eventually destructive contributions of various cavity modes over finite in-plane distances that limited us to using only a fraction of the cavity width in our example, this leads to significant constraints on the dimensions and relative placements of the FI and SC. However, Ref.~\cite{Zaytseva2020} reports out-of-plane critical fields of \si{nm}-thin Nb films of roughly $1$--$\SI{4}{T}$, while Ref.~\cite{Lin2020} reports out-of-plane coercivities in \si{nm}-thin Bi-doped YIG films of roughly $\SI{3e-4}{T}$. An aligning field can therefore be many orders of magnitude smaller than the SC critical field with appropriate material choices. One would then expect the effect of $\vec{B}_{\mathrm{ext}}$ on the SC to be negligible. However, we have not considered here the subsequent effect of the SC on the spatial distribution of $\vec{B}_{\mathrm{ext}}$, which was taken to be uniform across the FI.

Moreover, the Pearl length criterion, which greatly limits SC dimensions, can potentially be disregarded if the odd $\vec{p}$ symmetry of the anisotropy field~\eqref{eq:hkd} is broken by other means than a supercurrent. A candidate for this is taking into account spin--orbit coupling on the SC and subjecting it to a weak (non-pair breaking) magnetostatic field.

Furthermore, in our set-up, we have considered coupling to the quasiparticle excitations of the SC. This has partly been motivated by the prospect of using the FI to probe detailed spin and momentum information about the SC gap, which would require an extension of our present model. Another interesting avenue to explore is coupling directly to the gap by considering fluctuations from its mean-field value. This has been explored for an FI--SC bilayer, where the Higgs mode of the SC couples linearly to a spin exchange field~\cite{Lu2022}. This has a significant impact on the SC spin susceptibility in a bilayer set-up.

Despite coupling to the quasiparticles, we find that the anisotropy field magnitude nearly constant at low temperatures, and rapidly decreases to zero near the critical temperature. This can be understood from the fact that the symmetry-breaking supercurrent momentum enters the system Hamiltonian via the gap (cf. Eq.~\eqref{eq:HBCSInit}). Hence, when the gap vanishes, so does the quantity that breaks the symmetry. On the other hand, for temperatures substantially below $T_{c0}$, the gap varies little with temperature; the anisotropy field becomes close to constant, with a magnitude depending on the momentum associated with the inversion symmetry-breaking current $\vec{P}$.

In the normal state, the DC through the SC induces a surrounding magnetostatic field, by the Biot--Savart law. This differs from the response in the superconducting state by instead being appreciable above $T_{c0}$, and by its spatial distribution; for instance, the magnetostatic field cannot reverse the field direction as observed between Fig.~\ref{fig:Lsep1} and \ref{fig:Lsep2}.

Lastly, it is seen from Eq.~\eqref{eq:hkd} that the SC quasiparticle modes uniformly affect the anisotropy field in our current set-up, as the sum over fermion momenta $\vp$ can be factored out from the sum over photon momenta $\vq$. This limits the resolution of SC features in the anisotropy field, and by extension the FI. However, to higher order in the calculations, the quantity $G_{\varsigma \varsigma^\prime}^{q q^\prime}$ defined in Eq.~\eqref{eq:GCoeffBilEffCav} enters, with sums over fermion momenta $\vp$ and $\vp^\prime$ that are inseparable from the cavity momenta $\vq$ and $\vq^\prime$. This quantity is a candidate for extracting more features of the SC via the FI.

\begin{acknowledgments}
We acknowledge funding via the ``Outstanding Academic Fellows'' programme at NTNU, the Research Council of Norway Grant number 302315, as well as through its Centres of Excellence funding scheme, project number 262633, “QuSpin”.
\end{acknowledgments}

\appendix

\section{Integrating out the SC first} \label{app:SCfirst}
The order in which we integrate out the cavity and the SC is inconsequential. We show this here by integrating out the SC first, starting from the partition function~\eqref{eq:partitionFunc}.

We introduce the interaction matrix $G$ with elements
\begin{equation}
    G_{m m^\prime}^{p p^\prime} \equiv \frac{1}{\sqrt{\beta}} \sum_{q \varsigma} g_{\varsigma m m^\prime}^{q p p^\prime} ( a_{q \varsigma} + a_{-q \varsigma}^\dagger ),
\end{equation}
and furthermore the diagonal matrix $E$ with elements
\begin{equation}
    E_{m m^\prime}^{p p^\prime} \equiv E_{p m} \delta_{p p^\prime} \delta_{m m^\prime}.
\end{equation}
Hence the action involving the SC can be written as
\begin{equation}
    S^{\mathrm{SC}}_0 +  S^{\mathrm{cav}-SC}_{\mathrm{int}} = \sum_{p m} \sum_{p^\prime m^\prime} ( E +  G )_{m m^\prime}^{p p^\prime} \gamma_{p m}^\dagger \gamma_{p^\prime m^\prime}.
\end{equation}
The part of the partition function~\eqref{eq:partitionFunc} which depends on the SC is a Gaussian integral, and can now be written as \cite{Altland}
\begin{equation}
\begin{split} \label{eq:partitionSC}
    Z^{\mathrm{SC}} & \equiv \int \mathcal{D}[\gamma, \gamma^\dagger] \exp\left[ - \frac{1}{\hbar} \sum_{p m} \sum_{p^\prime m^\prime} ( E +  G )_{m m^\prime}^{p p^\prime} \gamma_{p m}^\dagger \gamma_{p^\prime m^\prime} \right] \\
    & \approx \exp \left[ \mathrm{Tr} \left[ E^{-1} G - E^{-1} G E^{-1} G / 2 \right] \right].
\end{split}
\end{equation}
In the last line, we neglected a factor $\exp \mathrm{Tr} \ln \left( \beta E/\hbar \right)$ that is constant with respect to the integration variables, and expanded another logarithm to second order in $|E^{-1} G|$. Hence, integrating out the SC to second order in the cav--SC coupling yields an effective action
\begin{equation}
\begin{split}
    S^{\mathrm{cav}}_1 \equiv & -\hbar \mathrm{Tr} \left[ E^{-1} G - E^{-1} G E^{-1} G / 2 \right] \\
    = & - \frac{\hbar}{\sqrt{\beta}}\sum_{q \varsigma} \sum_{p m} \frac{g_{\varsigma m m}^{q p p}}{E_{p m}} ( a_{q \varsigma} + a_{-q \varsigma}^\dagger ) \\
    & + \sum_{q \varsigma} \sum_{q^\prime \varsigma^\prime} G_{\varsigma \varsigma^\prime}^{q q^\prime} ( a_{q \varsigma} + a_{-q \varsigma}^\dagger ) ( a_{q^\prime \varsigma^\prime} + a_{-q^\prime \varsigma^\prime}^\dagger ),
\end{split}
\end{equation}
where we introduced the coefficient
\begin{equation} \label{eq:GCoeffBilEffCav}
    G_{\varsigma \varsigma^\prime}^{q q^\prime} \equiv \frac{\hbar}{2\beta} \sum_{p m} \sum_{p^\prime m^\prime} \frac{g_{\varsigma m m^\prime}^{q p p^\prime} g_{\varsigma^\prime m^\prime m}^{q^\prime p^\prime p}}{E_{p m} E_{p^\prime m^\prime}}.
\end{equation}

We now proceed to isolate the photonic terms and integrate out the cavity, i.e., we will perform the integral
\begin{equation}
\begin{split} \label{eq:partitionCav}
    Z^{\mathrm{cav}} \equiv \int \mathcal{D}[a, a^\dagger] e^{-S^{\mathrm{cav}}/\hbar},
\end{split}
\end{equation}
where the effective cavity action is
\begin{equation} \label{eq:SCavEff}
    S^{\mathrm{cav}} \equiv S^{\mathrm{cav}}_0 + S^{\mathrm{cav}}_1 + S^{\mathrm{FI-cav}}_{\mathrm{int}}.
\end{equation}
To this end, we introduce the current operator
\begin{equation} \label{eq:actionCurrent}
    J_{q \varsigma} \equiv -\sum_{k \p} G_{\p \varsigma}^{k q} ( \nu_{\p} \eta_{-k} + \nu_{\p}^* \eta_k^\dagger) + s_{q \varsigma},
\end{equation}
and perform a shift of integration variables
\begin{subequations}
\begin{align}
    a_{q \varsigma} & \rightarrow a_{q \varsigma} + J_{-q \varsigma}/\hbar \omega_q, \label{eq:aShift1} \\
    a_{q \varsigma}^\dagger & \rightarrow a_{q \varsigma}^\dagger + J_{q \varsigma} / \hbar \omega_q. \label{eq:aShift2}
\end{align}
\end{subequations}
The quantities $G_{\p \varsigma}^{k q}$ (to be distinguished from $G_{\varsigma \varsigma^\prime}^{q q^\prime}$) and $s_{q \varsigma}$ are coefficients of linear photon terms to be determined.

We now require that the shifts~\eqref{eq:aShift1}--\eqref{eq:aShift2} absorb the explicit linear photon terms in the action~\eqref{eq:SCavEff}, leaving only bilinear and constant terms in the shifted variables. This leads to self-consistency equations for $G_{\p \varsigma}^{k q}$ and $s_{q \varsigma}$. However, to second order in $|E^{-1} G|$, it can be shown that only the lowest-order expressions for $G_{\p \varsigma}^{k q}$ and $s_{q \varsigma}$ affect the anisotropy field to be extracted at the end, \cf Sec.~\ref{subsec:result_electric}. These are
\begin{align}
    G_{\p \varsigma}^{k q} & = g_{\p \varsigma}^{k q}, \\
    s_{q \varsigma} & = \frac{\hbar}{\sqrt{\beta}} \sum_{p m} \frac{g_{\varsigma m m}^{q p p}}{E_{p m}}.
\end{align}
Hence, the action~\eqref{eq:SCavEff} can be written as
\begin{equation}
    \begin{split} \label{eq:SCavEffNew}
        S^{\mathrm{cav}} = S^{\mathrm{cav}}_{\mathrm{bil}} + S^{\mathrm{cav}}_{\mathrm{con}}
    \end{split}
\end{equation}
where
\begin{widetext}
\begin{equation}
    \begin{split}
        S^{\mathrm{cav}}_{\mathrm{bil}} \equiv \sum_{q \varsigma} \hbar \omega_q a_{q \varsigma}^\dagger a_{q \varsigma} + \sum_{q \varsigma} \sum_{q^\prime \varsigma^\prime} G_{\varsigma \varsigma^\prime}^{q q^\prime} ( a_{q \varsigma} + a_{-q \varsigma}^\dagger ) ( a_{q^\prime \varsigma^\prime} + a_{-q^\prime \varsigma^\prime}^\dagger ),
    \end{split}
\end{equation}
\begin{equation}
    \begin{split}
        S^{\mathrm{cav}}_{\mathrm{con}} \equiv & \sum_{q \varsigma} \frac{J_{q \varsigma} J_{-q \varsigma}}{\hbar \omega_q} + \sum_{q \varsigma} \sum_{q^\prime \varsigma^\prime} G_{\varsigma \varsigma^\prime}^{q q^\prime} J_{-q \varsigma} J_{-q^\prime \varsigma^\prime} \left[ \frac{1}{\hbar\omega_q} + \frac{1}{\hbar\omega_{-q}} \right] \left[ \frac{1}{\hbar\omega_{q^\prime}} + \frac{1}{\hbar\omega_{-q^\prime}} \right].
    \end{split}
\end{equation}
\end{widetext}
$S^{\mathrm{cav}}_{\mathrm{bil}}$ contains all bilinear terms with respect to the shifted variables, and $S^{\mathrm{cav}}_{\mathrm{con}}$ all constant terms.

Returning to the integral~\eqref{eq:partitionCav}, by Eq.~\eqref{eq:SCavEffNew}, we now have
\begin{equation}
\begin{split} \label{eq:partitionCavCompletedSquare}
    Z^{\mathrm{cav}} = \int \mathcal{D}[a, a^\dagger] e^{-S^{\mathrm{cav}}/\hbar} = e^{-S^{\mathrm{cav}}_{\mathrm{con}}/\hbar} \int \mathcal{D}[a, a^\dagger] e^{-S^{\mathrm{cav}}_{\mathrm{bil}}/\hbar}.
\end{split}
\end{equation}
The integrand is now independent of magnons, and therefore inconsequential to the physics of the ferromagnetic insulator. We can therefore neglect the integral, leaving only the exponential prefactor. We are thus left with an effective FI partition function
\begin{equation}
    Z^{\mathrm{FI}} \equiv \int \mathcal{D}[\eta, \eta^\dagger] e^{-S^{\mathrm{FI}}/\hbar},
\end{equation}
where the effective FI action is
\begin{equation}
    S^{\mathrm{FI}} \equiv S^{\mathrm{FI}}_0 + S^{\mathrm{cav}}_{\mathrm{con}}.
\end{equation}
Neglecting magnon-independent terms, $S^{\mathrm{FI}}$ reads, after some rewriting,
\begin{widetext}
\begin{equation} \label{eq:SFIEff}
    S^{\mathrm{FI}} = \sum_{k} \hbar \lambda_k \eta_{k}^\dagger \eta_{k} + \sum_{k \p} \sum_{k^\prime \p^\prime} Q_{\p \p^\prime}^{k k^\prime} (\nu_{\p} \eta_{-k} + \nu_{\p}^* \eta_{k}^\dagger) (\nu_{\p^\prime} \eta_{-k^\prime} + \nu_{\p^\prime}^* \eta_{k^\prime}^\dagger) - g\mu_B\sum_{k\p} h_\p^k \cdot \sqrt{\frac{S}{2}} (\nu_\p \eta_{-k} + \nu_\p^* \eta_{k}^\dagger).
\end{equation}
Above, we introduced
\begin{equation}
    \begin{split} \label{eq:Qkkss}
        Q_{\p \p^\prime}^{k k^\prime} \equiv - \sum_{q \varsigma} \left[ \frac{ g_{\p \varsigma}^{k q} g_{\p^\prime \varsigma}^{k^\prime -q} }{\hbar \omega_q} + \sum_{q^\prime \varsigma^\prime} G_{\varsigma \varsigma^\prime}^{q q^\prime} \left[ \frac{1}{\hbar\omega_q} + \frac{1}{\hbar\omega_{-q}} \right] \left[ \frac{1}{\hbar\omega_{q^\prime}} + \frac{1}{\hbar\omega_{-q^\prime}} \right] g_{\p \varsigma}^{k q} g_{\p^\prime \varsigma^\prime}^{k^\prime q^\prime} \right],
    \end{split}
\end{equation}
\begin{equation}
    \begin{split} \label{eq:Pks}
        h^k_\p ={}& -\frac{\hbar}{g\mu_B} \sqrt{\frac{2}{S\beta}}\sum_{pm} \frac{V_{\p mm}^{kpp}}{E_{pm}},
    \end{split}
\end{equation}
which to leading order in the paramagnetic coupling are indeed the same as Eqs.~\eqref{eq:hkd} and \eqref{eq:Qkkdd}.
\end{widetext}

\section{SC at magnetic antinode}\label{subsec:results_magnetic}

\begin{figure}[h!tbp]
    \centering
    \includegraphics[width=1\linewidth]{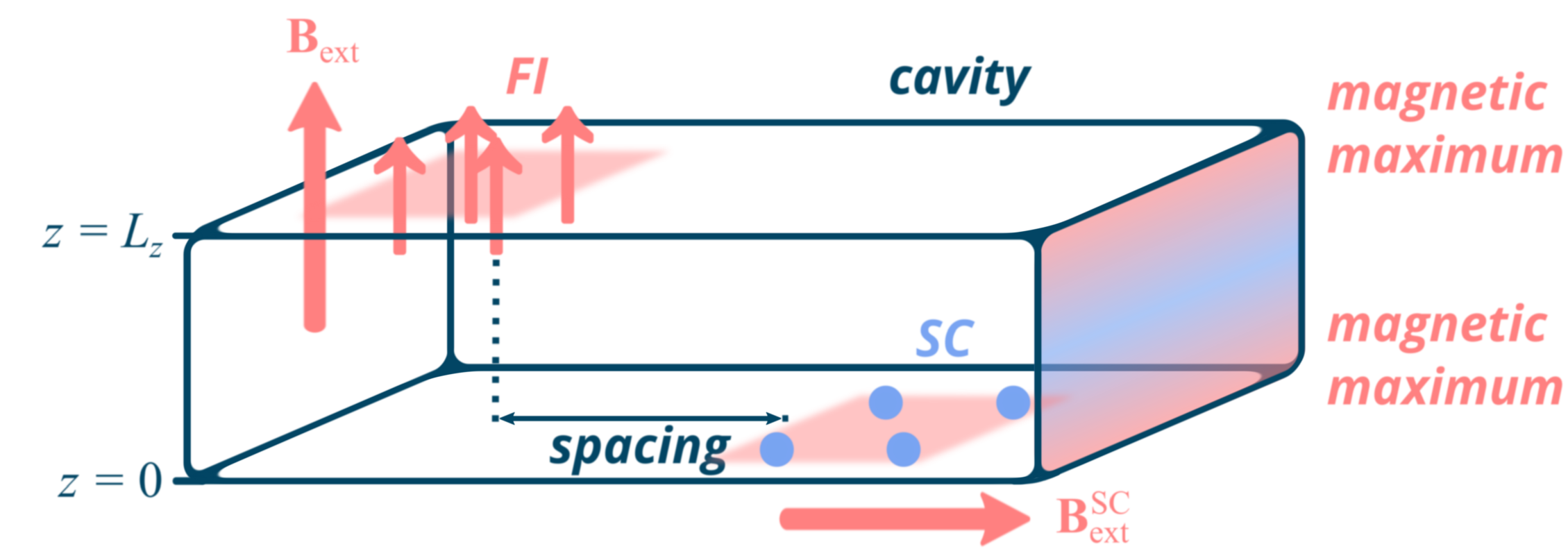}
    \caption{Illustration of the set-up with the SC placed at the magnetic antinode. The SC is subjected to an aligning external in-plane magnetic field $\vec{B}_{\mathrm{ext}}^{\mathrm{SC}}$. This set-up is otherwise identical to the one illustrated in Fig.~\ref{fig:set-up}.}
    \label{fig:set-up_magnetic}
\end{figure}

To compare our results for the FI-SC coupling with the SC placed at the electric antinode, we examine what happens when we place the superconductor at a magnetic maximum at $z \approx 0$, \cf \cref{fig:set-up_magnetic}. In this case the vector potential $\vec{A}$ points purely in the $z$ direction, and therefore does not couple to the SC via the paramagnetic coupling term used above. We therefore couple the SC to the cavity via the Zeeman coupling, and calculate the resulting anisotropy field across the FI. For the setup considered in the main text, it was necessary to break the inversion symmetry to get a finite anisotropy field, achieved, for instance, by applying a DC current. For the present setup, it is necessary to break the in-plane spin rotation symmetry, which can be achieved by applying an in-plane magnetic field to the SC. This becomes evident when considering the coupling between the cavity and SC. Placing the SC at $z\approx 0$, the cavity magnetic field is purely in-plane, pointing in the opposite direction to the field at $z=L_z$ [\cref{eq:Bcav}], resulting in a coupling term,
\begin{align}
    S_\mathrm{Zeeman} = \sum_{qpp'}\sum_{\sigma\sigma'} g_{\sigma\sigma'}^{qpp'}
    (a_{q1} + a_{-q1}^\dagger)c_{p\sigma}^\dagger c_{p'\sigma'},
\end{align}
with interaction matrix
\begin{align}
    g_{\sigma\sigma'}^{qpp'} ={}& \delta_{\On,\on-\on'} \nonumber\\*
    &\times\sqrt{\frac{\hbar \mu_B^2}{\epsilon \omega_{\vec{q}}V}} D_{\vec{p}-\vec{p}',\vec{q}}^{\mathrm{SC}} e^{i\vec{q}\cdot \vec{r}_0^{\mathrm{SC}}} i\sin\theta_{\vec{q}} (\bm{\sigma}\times \vec{q})_{\sigma\sigma'}\cdot \hat{e}_z.\label{eq:gcavSCmag}
\end{align}
This interaction alone would lead to a SC-cavity coupling that is off-diagonal in quasiparticle basis. The anisotropy field, corresponding to the diagram for $S_\mathrm{int}$ in \cref{fig:interaction_diagrams} with connected quasiparticle lines will therefore be exactly zero unless one breaks the spin-rotation symmetry by an in-plane magnetostatic field $\vec{B}_\mathrm{ext}^\mathrm{SC}$. The latter can for example be experimentally realized using external coils, as suggested for $\vec{B}_\mathrm{ext}$. In that case the quasiparticle bands are spin-split, resulting in the SC term
\begin{align}
    S_0^\mathrm{SC} = \sum_{pn} (-i \hbar \on + E_{\vp n})\gamma^\dagger_{pn}\gamma_{pn},
\end{align}
with the four quasiparticle bands
\begin{align}
    E_{\vp n}= (-1)^{\lfloor n/2 \rfloor} E_{\vp} + (-1)^n H,
\end{align}
with $E_{\vp} = \sqrt{\xi_{\vp}^2 + |\Delta_{\vp}|^2}$, $n \in [0, 1, 2, 3]$ and $H = |\mu_B \vec{B}_\mathrm{ext}^\mathrm{SC}|$. The bands are independent of in-plane direction of the field $\vec{B}_\mathrm{ext}^\mathrm{SC}$, with the directional dependence entering through the coupling between the quasiparticles and the cavity photons,
\begin{align}
    S_\mathrm{int}^\mathrm{SC-cav} ={}& \frac{1}{2\sqrt{\beta}}\sum_{qpp}\sum_{nn'} g_{nn'}^{qpp'}(a_{q1} + a_{-q1}^\dagger) \gamma_{pn}^\dagger \gamma_{p'n'},
\end{align}
where we have defined the interaction matrix in the Bogoliubov quasiparticle basis
\begin{widetext}
\begin{align}
    g_{nn'}^{qpp'} ={}& -\frac{1}{2}g_{ \su\sd}^{qpp'}e^{i\phi}\begin{pmatrix}
    [u_p^\dagger u_{p'} + v_{p} v_{p'}^\dagger ][\sigma_z + i \sigma_y] & [u_p^\dagger v_{p'} - v_p u_{p'}^\dagger][\sigma_0 -\sigma_x]\\
    [v_p^\dagger u_{p'} - u_p v_{p'}^\dagger][\sigma_0 + \sigma_x] & [v_{p}^\dagger v_{p'} + u_p u_{p'}^\dagger][\sigma_z - i \sigma_y]
    \end{pmatrix}_{nn'}\nonumber\\*
    &- \frac{1}{2}g_{\sd\su}^{qpp'}e^{-i\phi}\begin{pmatrix}
    [u_p^\dagger u_{p'} + v_p v_{p'}^\dagger][\sigma_z - i \sigma_y] & [u_p^\dagger v_{p'} - v_p u_{p'}^\dagger][\sigma_0 +\sigma_x]\\
    [v_p^\dagger u_{p'} - u_p v_{p'}^\dagger][\sigma_0 - \sigma_x] & [v_p^\dagger v_{p'} + u_p u_{p'}^\dagger][\sigma_z + i \sigma_y]
    \end{pmatrix}_{nn'}\label{eq:gsnnqppmag},
\end{align}
\end{widetext}
where $\sigma_0$ is the $2\times2$ identity matrix, and $\phi$ is the angle of the in-plane field relative to the $x$ axis. We have also defined the functions
\begin{subequations}
\begin{align}
    u_{\vp} ={}& e^{i\theta_{\vp}}\sqrt{\frac{1}{2}\l(1 + \frac{\xi_{\vp}}{E_{\vp}}\r)},\\
    v_{\vp} ={}& e^{i\theta_{\vp}}\sqrt{\frac{1}{2}\l(1 - \frac{\xi_{\vp}}{E_{\vp}}\r)},
\end{align}
\end{subequations}
which satisfy $|u_{\vp}|^2 + |v_{\vp}^2| = 1$. Here $2\theta_{\vp}$ is the phase of the order parameter.

Following the same procedure of integrating out the cavity photons and quasiparticles in the SC, we get an expression identical to \cref{eq:SFIeff}, with the only change coming in the anisotropy field, which is now defined as 
\begin{align}
    h_{\p}^k \equiv -\frac{\hbar}{\sqrt{2S \beta}g\mu_B}\sum_{pn} \frac{V_{\p nn}^{kpp}}{E_{pn}}, \label{eq:hskmag}
\end{align}
with
\begin{align}
    V_{\p nn'}^{kpp'} = \sum_{q}g_{\p 1}^{kq}g_{ nn'}^{-qpp'}\l[\frac{1}{\hbar\omega_q} + \frac{1}{\hbar\omega_{-q}}\r]\label{eq:Vmag}.
\end{align}
The additional factor of $1/2$ in the definition of $h_{\p}^k$ is due to the field integral resulting in the Pfaffian of the antisymmetrized Green's function in this case, which is the square root of the determinant \cite{Wegner2016}. The reason for this is the necessity of an expanded Nambu spinor, which contains both creation and annihilation operators of both types of quasiparticles when including an in-plane field \cite{Krohg2018}.

\begin{figure}[h!tbp]
    \includegraphics{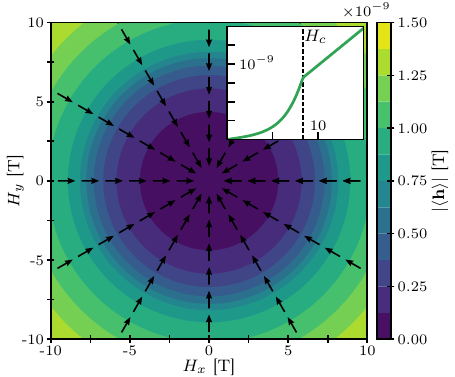}
    \caption{Absolute value (contour plot) and direction (arrows) of the averaged anisotropy field as a function of applied field strength and direction. The anisotropy field points opposite the applied field over the SC, following a $\cos\phi$ and $\sin\phi$ dependence for the $x$ and $y$ component respectively. The inset shows the absolute value of the in-plane projection as a function of the field strength. The temperature is set to $T = 0.5 T_{c0}$. The cavity dimensions are $L_x = L_y = L =\SI{10}{cm}$ and $L_z = \SI{1}{mm}$, and the FI and SC have sides of length $0.001L$ in the $x$ and $y$ directions, and are placed at the center of the cavity. The thickness of the SC is $d_{\mathrm{SC}} = \SI{10}{\nano\meter}$.}
    \label{fig:h_vs_phi}
\end{figure}

Inserting \cref{eq:Vmag,eq:gsnnqppmag} into \cref{eq:hskmag} and performing the sum over fermionic Matsubara frequencies \cite{Altland}, we get
\begin{align}
    h_{\p}^k ={}& \frac{\sqrt{\beta}\delta_{\Om0}}{\sqrt{2S}g\mu_B} \sum_{\vq\vp}\frac{g_{\p}^{\vk\vq}}{\hbar\omega_{\vq}}[g_{\su\sd}^{-\vq\vp\vp}e^{i\phi} + g_{ \sd\su}^{-\vq\vp\vp}e^{-i\phi}]\nonumber\\*
    &\times\l[\tanh\frac{\beta(E_{\vp}+H)}{2\hbar} - \tanh\frac{\beta(E_{\vp}-H)}{2\hbar} \r],
\end{align}
where we have used the fact that $\omega_{\vq}$ is even in $\vq$. Here it is clear that the anisotropy field is exactly zero when the in-plane field is zero, since the last two terms exactly cancel in that case. Moreover, since the anisotropy field is independent of the frequency $\Om$, we define the time-independent anisotropy field $h_{\p}^{\vk} = \sum_{\Om}h_\p^k e^{-i\Om\tau}/\sqrt{\beta}$. Inserting the expressions for $g_{\p}^{\vk\vq}$ and $g_{\sigma\sigma'}^{-\vq\vp\vp}$ from \cref{eq:magnoncoupling2,eq:gcavSCmag} we get
\begin{widetext}
\begin{align}
    h_{\p}^{\vk} ={}& -\frac{\mu_B\sqrt{N_\mathrm{FI}}}{\epsilon V} \sum_{\vq} e^{i\vec{q}\cdot (\vec{r}_0^{\mathrm{FI}} - \vec{r}_0^{\mathrm{SC}})} \frac{D_{\vk \vq}^\mathrm{FI}D_{\vec{0}\vq}^{\mathrm{SC}*}\sin^2\theta_{\vq}}{\omega_{\vq}^2}q_{\bar{\p}}\nu_{\p}^2 [q_y \cos\phi  - q_x \sin\phi]\sum_{\vp} \l[\tanh\frac{\beta(E_{\vp} + H)}{2\hbar} - \tanh\frac{\beta(E_{\vp} - H)}{2\hbar} \r].\label{eq:anisotropyfieldmag}
\end{align}

We focus on the anisotropy field averaged across the FI, $ \langle h_\p \rangle = \sum_i h_\p(\vec{r}_i,\tau)/N_\mathrm{FI} = \sum_i \sum_{\vk}  h_{\p}^{\vk} e^{i \vk \cdot \vec{r}_i}/N_{\mathrm{FI}}^{3/2} = h_{\p}^{\vec{0}}/\sqrt{N_{\mathrm{FI}}}$ (\cf Eq.~\eqref{eq:hdhkd}), rewrite the first sum such that it becomes dimensionless, and transform the second sum into an integral using a free electron gas dispersion $\xi_\vec{k} = \hbar^2\vec{p}^2/2m - \mu$. Assuming cavity dimensions $L_x = L_y = L$ and an $s$-wave gap, we get
\begin{align}
    \langle h_\p \rangle ={}& -\frac{\mu_B V_\mathrm{SC} (m\Delta_0)^{3/2}}{\sqrt{2}\pi^2\hbar^3\epsilon c^2V} \sum_{\vq} e^{i\vec{q}\cdot (\vec{r}_0^{\mathrm{FI}} - \vec{r}_0^{\mathrm{SC}})} D_{0\vq}^\mathrm{FI}D_{0\vq}^{\mathrm{SC}*}\frac{\ell_{\bar{\p}}\nu_\p^2[\ell_y\cos\phi - \ell_x\sin\phi][\ell_x^2 + \ell_y^2]}{\l[\ell_x^2 + \ell_y^2 + \l(\frac{L}{2L_z}\r)^2\r]^2}\nonumber\\*
    &\times \int\limits_{-\mu/\Delta_0}^{\xi_\mathrm{max}/\Delta_0} dx \sqrt{x + \frac{\mu}{\Delta_0}}\bigg[\tanh \frac{1.764T_c\l(\sqrt{x^2 + |\Delta/\Delta_0|^2} + H/\Delta_0\r)}{2T} - \tanh\frac{1.764T_c\l(\sqrt{x^2 + |\Delta/\Delta_0|^2} - H/\Delta_0\r)}{2T}\bigg]. \label{eq:avganisotropyfieldmag}
\end{align}
\end{widetext}
Here $V_\mathrm{SC}$ and $\Delta_0$ are the volume and zero temperature gap of the superconductor, respectively, and $m$ the electron mass. $\ell_x$ and $\ell_y$ are integer indexes corresponding to cavity momentum $\vq$. From the above expression we expect terms even in $\ell_\p$ to dominate, resulting in the anisotropy field and expectation values of the in-plane spin components to have a $\phi$ dependence given by $h_x^\vk \sim \langle S_{ix} \rangle \propto - \cos\phi$ and $h_y^\vk \sim \langle S_{iy} \rangle \propto -\sin \phi$. This is in good agreement with numerical solutions of \cref{eq:avganisotropyfieldmag} in an arbitrary practical example, as shown in \cref{fig:h_vs_phi}. The results were obtained using the Python libraries \texttt{NumPy} and \texttt{Matplotlib}, and sub-package \texttt{scipy.integrate}. Notice, however, that the magnitude of the anisotropy field is very small, on the order of $\SI{E-9}{\tesla}$. This is several orders of magnitude smaller than the previously considered setup, and we do not expect this to be a measurable effect. Here we have neglected the effect of an in-plane finite separation between the SC and FI by placing them directly above each other. A finite separation would further reduce the anisotropy field.

At zero temperature the two hyperbolic tangent functions in \cref{eq:avganisotropyfieldmag} are always equal to one, as long as $H < \Delta_0$. Since the field must be below the critical field $H_{c0} = \Delta_0/\sqrt{2}$ in the superconducting state, the two terms in the integral always cancel exactly at zero temperature. On the other hand, in the case of temperatures just above the critical temperature, $T \gtrsim T_c$, and $\mu,\xi_\mathrm{max} > H$, we get the analytical result $4H\sqrt{\mu}/\Delta_0^{3/2}$ for the integral, assuming that the main contribution to the integral comes from energies close to the Fermi level. Hence we expect the anisotropy field to increase from zero to the normal state value as temperature increases towards $T_c$, and that $\langle h_\p \rangle$ increases linearly with applied field in the normal state. This is found to be in good agreement with numerical results, see the inset in \cref{fig:h_vs_phi} for $|\vec{H}| > H_c$. In the numerical calculations we have assumed $\mu,\xi_\mathrm{max} \gg \Delta_0$, and that the gap's dependence on temperature and applied field is described by \cref{eq:gapInterpolation} multiplied with $\sqrt{1-(H/H_{c})^2}$ \cite{Gross1986,Douglass1961}, and the critical field depends on temperature as $H_c = H_{c0}[1 - (T/T_{c0})^2]$ \cite{Tinkham}, where $T_{c0}$ is the critical temperature for zero field.  Below the critical temperature and field, the field-dependence of the anisotropy field is more complicated due to the additional effect of reducing the superconducting gap, see inset in \cref{fig:h_vs_phi}. The difference in temperature and applied field-dependence of the anisotropy field between the normal and superconducting state could therefore in principle be a way of detecting the onset of superconductivity without directly probing the superconductor, though the anisotropy field calculated in this arbitrary example is too small to be detectable.

\section{Linear terms as an anisotropy field}\label{app:eta_shift}
In this appendix, we take a closer look at the interpretation of the linear magnon terms as interactions with an effective anisotropy field. Consider an FI in an inhomogeneous applied field,
\begin{align}
	\mathcal{H} = -J\sum_{\l<i,j\r>} \vec{S}_i\cdot\vec{S}_j -\sum_i \vec{H}_i\cdot \vec{S}_i.
\end{align}
Above, $\vec{H}_i = (H_i^x, H_i^y, H^z)$ is the inhomogeneous external field, with $H^z$ assumed homogeneous and much larger than $H_i^x, H_i^y$. We therefore assume ordering in the $z$ direction when performing the Holstein--Primakoff transformation, resulting in the Fourier-transformed Hamiltonian
\begin{align}
	\mathcal{H} = E_0 + \sum_{\vk} \l[\hbar\lambda_{\vk} \eta_{\vk}^\dagger \eta_{\vk} -h_{\vk}\eta_{\vk}^\dagger - h_{\vk}^*\eta_{\vk}\r].
\end{align}
Here $\hbar\lambda_{\vk}$ is the dispersion defined in \cref{eq:magnondispersion}, the classical ground state energy is
\begin{align}
	E_0 = -\hbar S N_\mathrm{FI}\l[J\hbar S N_\delta + H_z\r],
\end{align}
and the momentum-dependent in-plane magnetic energy
\begin{align}
	h_{\vk} ={}& \sqrt{\frac{S}{2N_\mathrm{FI}}}\hbar\sum_i (H_i^x + i H_i^y) e^{ -i\vk\cdot \vec{r}_i}.
\end{align}

Since the applied field has in-plane components, the $z$ direction is not the exact ordering direction in the ground state, leading to a non-diagonal Hamiltonian with linear terms. To get rid of these terms, we translate the fields according to
\begin{equation} \label{eq:etakShift}
\begin{aligned}
    \eta_{\vec{k}} \to \eta_{\vec{k}} + t_{\vec{k}},\\
    \eta_{\vec{k}}^\dagger \to \eta_{\vec{k}}^\dagger + t_{\vec{k}}^*,
\end{aligned}
\end{equation}
and require that linear terms cancel. Translating the fields leads to the Hamiltonian
\begin{align}
	\mathcal{H} \to {}& E_0 + \sum_{\vk} \Big\{\hbar\lambda_{\vk} \eta_{\vk}^\dagger \eta_{\vk} + [\hbar\lambda_{\vk} t_{\vk} - h_{\vk}]\eta_{\vk}^\dagger\nonumber\\*
	&+ [\hbar\lambda_{\vk} t_{\vk}^* - h_{\vk}^*]\eta_{\vk} + \hbar\lambda_{\vk} t_{\vk}^*t_{\vk} -h_{\vk} t_{\vk}^* - h_{\vk}^*t_{\vk}\Big\},
\end{align}
and we therefore require
\begin{align}
	t_{\vk} ={}& \frac{h_{\vk}}{\hbar\lambda_{\vk}}.
\end{align}
The resulting diagonal Hamiltonian is
\begin{align}
	\mathcal{H} = E_0 + \sum_{\vk} [\hbar\lambda_{\vk} \eta_{\vk}^\dagger \eta_{\vk} - \hbar\lambda_{\vk} t_{\vk}^* t_{\vk}].
\end{align}

The last term in the above equation results in a renormalization of the classical ground state,
\begin{align}
	E_0 \to{}& E_0 - \sum_{\vk}\hbar\lambda_{\vk} t_{\vk}^*t_{\vk}\nonumber\\*
	={}& E_0 - \sum_{i,j,\vk}\frac{S\hbar^2(H_i^x + i H_i^y)(H_j^x - i H_j^y)^2e^{i\vk\cdot(\vec{r}_j - \vec{r}_i)}}{2N_\mathrm{FI}\hbar\lambda_{\vk}}.
\end{align}
In the case of constant in-plane components, this simplifies to
\begin{align}
    E_0 ={}& -\hbar S N_\mathrm{FI}\l\{J\hbar S N_\delta + \l[H^z + \frac{(H^x)^2 + (H^y)^2}{2H^z}\r]\r\}\nonumber\\*
    \approx{}& -\hbar S N_\mathrm{FI}\l[J\hbar S N_\delta + |\vec{H}|\r],
\end{align}
where the approximation in the last line is valid in the limit $|H^x|,|H^y| \ll |H^z|$. This is as expected, since the classical ground state is generally oriented along $\vec{H}$, not $H^z$. The translation of magnon operators in \cref{eq:etakShift} can therefore be understood as a local rotation of the spin ordering ansatz due to small inhomogeneous in-plane fields, valid in the limit $|H_i^{x,y}| \ll |H^z|$.

\end{document}